\documentclass[11pt]{article}
\usepackage{amsmath, amsfonts, setspace, url, hyperref, verbatim, xcolor, cite, graphicx, todonotes, cancel, slashed, mathtools}
\usepackage[T1]{fontenc}
\usepackage[utf8]{inputenc}
\usepackage[a4paper,top=2cm,left=2cm,right=2cm,bottom=2cm]{geometry}
\usepackage{tikz}

\newcommand{\dd}{\text{d}}

\newcommand{\fM}{\mathcal{M}}
\newcommand{\fN}{\mathcal{N}}
\newcommand{\fP}{\mathcal{P}}
\newcommand{\fQ}{\mathcal{Q}}

\newcommand{\fA}{\mathcal{A}}
\newcommand{\fB}{\mathcal{B}}
\newcommand{\fC}{\mathcal{C}}
\newcommand{\fD}{\mathcal{D}}
\newcommand{\fE}{\mathcal{E}}
\newcommand{\fF}{\mathcal{F}}

\newcommand{\be}{\begin{equation}}
\newcommand{\ee}{\end{equation}}
\numberwithin{equation}{section}

\newcommand{\gM}{\mathcal{M}}
\newcommand{\cH}{\mathcal{H}}

\newcommand{\Cdef}{\mathbf{C}}
\newcommand{\Fdef}{\mathbf{F}}

\newcommand{\Aa}{\mathcal{A}}
\newcommand{\Ab}{\mathcal{B}}
\newcommand{\Ac}{\mathcal{C}}

\newcommand{\Fa}{\mathcal{F}}
\newcommand{\Fb}{\mathcal{H}}
\newcommand{\Fc}{\mathcal{J}}

\newcommand{\Gfour}{\mathrm{SL}(5)}

\newcommand{\Gsix}{E_{6(6)}}

\newcommand{\Edd}{E_{d(d)}}

\newcommand{\Hfour}{\mathrm{SO}(5)}

\definecolor{vub}{RGB}{0,52,154}
\definecolor{vubo}{RGB}{255,102,0}
\definecolor{redd}{RGB}{255,40,40}
\definecolor{r}{RGB}{228,32,20}
\definecolor{o}{RGB}{238,69,4}
\definecolor{y}{RGB}{253,228,1}
\definecolor{g}{RGB}{108,160,0}
\definecolor{b}{RGB}{0,162,203}
\definecolor{i}{RGB}{120,42,117}

\usepackage[utf8]{inputenc}
\usepackage[framemethod=tikz]{mdframed}
\usetikzlibrary{calc}
\usepackage{chngcntr}

\counterwithin{ex}{section}

\makeatletter
\def\mdf@@mynote{}
\define@key{mdf}{mynote}{\def\mdf@@mynote{#1}}

\mdfdefinestyle{mystate}{
skipabove={1\baselineskip},
linewidth=0.5pt,
innertopmargin=1\baselineskip,
roundcorner=10pt,
backgroundcolor=black!0,
linecolor=black,
singleextra={
  \node[xshift=10pt,draw=black,fill=white,rounded corners,anchor=west] at (P-|O) %
  {\strut{\itshape  }\ifdefempty{\mdf@@mynote}{}{\itshape\bfseries \mdf@@mynote}};
},
firstextra={
  \node[xshift=10pt,draw=black,fill=white,rounded corners,anchor=west] at (P-|O) %
  {\strut{\itshape  }\ifdefempty{\mdf@@mynote}{}{\itshape\bfseries \mdf@@mynote}};
}
}

\mdfdefinestyle{mystateb}{
skipabove={1\baselineskip},
linewidth=1pt,
innertopmargin=1\baselineskip,
roundcorner=10pt,
backgroundcolor=white!5,
linecolor=vub!50,
singleextra={
  \node[xshift=10pt,thick,draw=vub!50,fill=b!0,rounded corners,anchor=west] at (P-|O) %
  {\strut{\itshape  }\ifdefempty{\mdf@@mynote}{}{\bf\mdf@@mynote}};
},
firstextra={
  \node[xshift=10pt,thick,draw=vub!50,fill=b!0,rounded corners,anchor=west] at (P-|O) %
  {\strut{\itshape  }\ifdefempty{\mdf@@mynote}{}{\bf\mdf@@mynote}};
}
}

\mdfdefinestyle{mystater}{
skipabove={1\baselineskip},
linewidth=1pt,
innertopmargin=1\baselineskip,
roundcorner=10pt,
backgroundcolor=white!5,
linecolor=r!50,
singleextra={
  \node[xshift=10pt,thick,draw=r!50,fill=b!0,rounded corners,anchor=west] at (P-|O) %
  {\strut{\itshape  }\ifdefempty{\mdf@@mynote}{}{\bf\mdf@@mynote}};
},
firstextra={
  \node[xshift=10pt,thick,draw=r!50,fill=b!0,rounded corners,anchor=west] at (P-|O) %
  {\strut{\itshape  }\ifdefempty{\mdf@@mynote}{}{\bf\mdf@@mynote}};
}
}

\mdfdefinestyle{mystateg}{
skipabove={1\baselineskip},
linewidth=1pt,
innertopmargin=1\baselineskip,
roundcorner=10pt,
backgroundcolor=white!5,
linecolor=g!50,
singleextra={
  \node[xshift=10pt,thick,draw=g!50,fill=g!0,rounded corners,anchor=west] at (P-|O) %
  {\strut{\itshape  }\ifdefempty{\mdf@@mynote}{}{\bf\mdf@@mynote}};
},
firstextra={
  \node[xshift=10pt,thick,draw=g!50,fill=g!0,rounded corners,anchor=west] at (P-|O) %
  {\strut{\itshape  }\ifdefempty{\mdf@@mynote}{}{\bf\mdf@@mynote}};
}
}

\newmdenv[style=mystate,nobreak=true]{state}
\newmdenv[style=mystater,nobreak=true]{stater}
\newmdenv[style=mystateg,nobreak=true]{stateg}
\newmdenv[style=mystateb,nobreak=true]{stateb}

\makeatother

\makeatother

\newcommand{\rpm}{-}
\newcommand{\rmp}{+}

\newcommand{\ii}{\mathsf{i}}
\newcommand{\jj}{\mathsf{j}}
\newcommand{\kk}{\mathsf{k}}
\renewcommand{\ll}{\mathsf{l}}
\newcommand{\mm}{\mathsf{m}}
\newcommand{\nn}{\mathsf{n}}
\newcommand{\pp}{\mathsf{p}}
\newcommand{\trivector}{\pi}
\newlength{\bibitemsep}
\newlength{\bibparskip}
\let\oldthebibliography\thebibliography
\renewcommand\thebibliography[1]{%
\oldthebibliography{#1}%
\setlength{\parskip}{\bibitemsep}%
\setlength{\itemsep}{\bibparskip}%
}
\begin{document}

\vfill

\begin{center}
	\baselineskip=16pt  
	
	{\Large \bf  
	Generalised U-dual solutions in supergravity
	}
	\vskip 2em
	{\large \bf  Chris D. A. Blair, Sofia Zhidkova}
	\vskip 0.6em
	{\it  
			Theoretische Natuurkunde, Vrije Universiteit Brussel, and the International Solvay Institutes, \\ Pleinlaan 2, B-1050 Brussels, Belgium 
			\\ {\tt Christopher.Blair@vub.be}, {\tt Sofia.Zhidkova@vub.be}
	}
	\vskip 2cm 
\end{center}

\begin{abstract}
We discuss the notion of generalised U-duality as a solution generating technique in supergravity. We demonstrate a method to take solutions of type IIA supergravity on a 3-sphere, with NSNS flux, to new solutions of 11-dimensional supergravity, using exceptional geometry techniques. These new solutions are characterised by an underlying 3-algebra structure, and generalise features of solutions obtained by non-abelian T-duality, which involve an underlying ordinary Lie algebra. We focus on an example where we start with the pp-F1-NS5 solution in type IIA supergravity. We discuss the properties of our resulting new solution, including the possibility of viewing it globally as a U-fold, and its M2 and M5 brane charges. In the extremal case, the new solution admits an AdS${}_3$ limit, which falls into a recently constructed class of M-theory AdS$_{3}$ backgrounds -- this provides a global completion of our solution with a well-defined holographic dual, similar to the global completions of non-abelian T-dual solutions. Our full solution is a 6-vector deformation of this AdS${}_3$ limit. We also explicitly solve the Killing spinor equation in the AdS${}_3$ limit, finding a $\frac{1}{2}$-BPS solution.
\end{abstract}

\tableofcontents

\section{Introduction}

This paper illustrates a method to take solutions of type IIA supergravity on a three-sphere, with NSNS flux, to new solutions of 11-dimensional supergravity on a four-dimensional space with particular properties.
Principal amongst these properties is that the geometry of this space is secretly controlled by an underlying algebraic structure incorporating the structure constants of a three-algebra symmetry.
This structure generalises that found in solutions generated by non-abelian T-duality, which produces geometries controlled by an underlying Lie algebra symmetry.
We focus on an example where we start with the F1-NS5 near horizon solution of type IIA supergravity, and construct a new 11-dimensional solution involving M2-M5-M5' charges.

The context for our work is the question of how to formulate and use \emph{generalised dualities} in M-theory. 
The classic formulation of a string or M-theory duality is in terms of an equivalence between theory 1 on space $X_1$ and theory 2 on space $X_2$.
Conventional (abelian) T-duality corresponds to the case when theory 1 is type IIA string theory, theory 2 is type IIB string theory, and $X_1$ and $X_2$ are circles of inverse radius.
U-duality can be stated as an equivalence between M-theory on dual $d$-dimensional tori, or type II theory on $(d-1)$-dimensional tori.

In supergravity, these dualities can be rephrased as expressing the fact that a dimensional reduction or consistent truncation of supergravity 1 on $X_1$ gives the same lower-dimensional theory as a reduction of supergravity 2 on $X_2$.
This allows duality to be used as a solution generating technique, where solutions of supergravity 1 of the form $M \times X_1$ can be mapped to solutions of supergravity 2 of the form $M \times X_2$, by reducing and uplifting.

Generalised T- and U-duality extend this notion of duality to special classes of dual spaces $X_1$ and $X_2$, which are not tori.
At a minimum, this is a solution generating method: given a supergravity solution meeting particular conditions, a generalised duality will produce a second supergravity solution related in a particular manner to the first. 
Whether this extends to a genuine duality of the full (quantum) string or M-theory is far from guaranteed, even in T-duality examples where worldsheet methods can be used to formulate aspects of the duality.
However, these techniques have proven their value in supergravity alone as a source of new solutions with applications to holography, integrability and other areas (see \cite{Thompson:2019ipl} for a review and further references).
It is perhaps also worth remembering that what is now known as U-duality first appeared -- almost accidentally -- in supergravity \cite{Cremmer:1978ds}, long before the idea of M-theory was developed \cite{Hull:1994ys,Witten:1995ex}.

The most well-appreciated generalisation of T- or U-duality is non-abelian T-duality (NATD) \cite{delaOssa:1992vci}.
This has a worldsheet derivation, at least for the transformation of the NSNS sector fields. 
The basic structure of this duality is that it takes a space with non-abelian isometries, for example a group manifold, to a space with fewer isometries. The dual solution is characterised by an underlying algebraic structure controlled by `dual' structure constants $\tilde f^{ab}{}_c \neq 0$ inherited from the Lie algebra of the original non-abelian symmetry.

Unlike abelian T-duality, the worldsheet path integral derivation of the dual background does not lead to global information, in particular about the range or periodicity of the dual coordinates \cite{Giveon:1993ai}.
It is however possible to find various arguments to globally `complete' the supergravity solution. 
For instance, combined with the correct transformations for the RR sector \cite{Sfetsos:2010uq}, non-abelian T-duality has been extensively applied to generate AdS solutions with interesting CFT duals.
A common approach for NATD solutions with an AdS factor is to find a holographic completion by embedding the NATD solution into a supergravity solution with a well-defined holographic interpretation, usually in terms of a quiver field theory stemming from an underlying Hanany-Witten brane configuration \cite{Lozano:2016kum}.
Alternatively, as pointed out in \cite{Fernandez-Melgarejo:2017oyu,Bugden:2019vlj}, non-abelian T-dual solutions could be viewed globally as T-folds. 

Both abelian and non-abelian T-duality are special cases of Poisson-Lie T-duality \cite{Klimcik:1995dy,Klimcik:1995ux}. 
This applies to $d$-dimensional backgrounds which may in general lack isometries, but which geometrically encode data associated to a $2d$-dimensional Lie algebra called the Drinfeld double. 
This can be made manifest by adopting a generalised geometric (or double field theory) description \cite{Hassler:2017yza,Demulder:2018lmj}. 
For backgrounds admitting Poisson-Lie T-duality there exists a \emph{generalised parallelisation} \cite{Grana:2008yw,Lee:2014mla} 
providing a consistent truncation to a lower dimensional gauged supergravity. 
In general, two inequivalent higher-dimensional solutions admitting consistent truncations to the same lower dimensional theory can be viewed as dual in the sense we are considering.
(Indeed, NATD was expressed in terms of consistent truncations \cite{Itsios:2012dc} some years prior to its doubled geometry formulation \cite{Hassler:2017yza,Demulder:2018lmj,Sakatani:2019jgu,Catal-Ozer:2019hxw}).

The generalised geometry approach opens the door to the study of new variants of U-duality, by using 
the exceptional generalised geometry (or exceptional field theory) description of 11-dimensional supergravity.
This led to the proposals for Poisson-Lie U-duality and an associated `exceptional Drinfeld algebra' (EDA) introduced in \cite{Sakatani:2019zrs,Malek:2019xrf} and further studied from a variety of angles in \cite{Sakatani:2020iad,Blair:2020ndg,Malek:2020hpo,Sakatani:2020wah,Musaev:2020nrt,Bugden:2021wxg,Sakatani:2021eqo}. 

Whereas the Drinfeld double naturally encodes a pair of ordinary Lie subalgebras, the content of the EDA is more exotic. 
The EDA itself is generically a Leibniz rather than a Lie algebra. 
For M-theory backgrounds, the structure constants of the EDA are assembled from those of a Lie algebra $f_{ab}{}^c$ and a `dual' 3-algebra with structure constants $\tilde f^{abc}{}_d$ (as well as other $n$-algebra structure constants if the dimension of the algebra is large enough).

In our previous paper \cite{Blair:2020ndg}, cases where $\tilde f^{abc}{}_d \neq 0$ but $f_{ab}{}^c = 0$ were studied. 
These should underlie backgrounds (termed `three-algebra geometries' in \cite{Blair:2020ndg}) analogous to those which are generated by non-abelian T-duality. 
A particularly simple example is the Euclidean 3-algebra in four-dimensions, $\tilde f^{abc}{}_d \sim \epsilon^{abc}{}_d$.
The EDA in this case is the Lie algebra $\mathrm{CSO}(4,0,1)$, and the generalised geometry construction gives a consistent truncation to seven-dimensional $\mathrm{CSO}(4,0,1)$ gauged supergravity.
An alternative consistent truncation in this case is provided by type IIA on $\text{S}^3$ with NSNS flux \cite{Cvetic:2000ah}. 
This gives a solution generating mechanism, whereby type IIA solutions of this form can be consistently truncated to solutions of the seven-dimensional $\mathrm{CSO}(4,0,1)$ gauged supergravity, and then uplifted to new solutions of 11-dimensional supergravity using the generalised geometric formulation of \cite{Sakatani:2019zrs,Malek:2019xrf,Blair:2020ndg}.

In this paper, we apply this logic to produce a new 11-dimensional solution starting with a non-extremal pp-F1-NS5 solution of type IIA, after taking the five-brane near horizon limit.
Our new 11-dimensional solution has the following properties:
\begin{itemize}
\item Just as for non-abelian T-duality, the global properties of the new solution are a priori unknown. It can be described using a non-geometric generalised frame involving a trivector linear in the new four-dimensional dual coordinates, and so one possible global interpretation is as a U-fold. {\itshape (See section \ref{Ufold}.)}
\item The new solution can be viewed as carrying M2 and M5 brane charges. {\itshape (See section \ref{solnsph}.)}
\item In the extremal case, it admits a limit in which it becomes AdS${}_3 \times \text{S}^3 \times \text{T}^4$ foliated over an interval. This solution fits into the general class of M-theory AdS${}_3$ solutions derived in \cite{Lozano:2020bxo}. These solutions are directly inspired by solutions generated by non-abelian T-duality, and provide a global completion of our solution (in this AdS limit), with a known holographic dual and brane interpretation. This is exactly analogous to NATD solutions. {\itshape (See section \ref{adslimit}.)}
\item The full extremal solution can be viewed as a deformation of the AdS${}_3$ limit generated by a six-vector deformation parameter valued in $\Gsix$. 
This deformation is inherited from an $SO(2,2)$ T-duality-valued bivector deformation of the extremal F1-NS5 near horizon solution, which describes the interpolation from the AdS${}_3$ near horizon region to an asymptotic linear dilaton spacetime. In that case, the deformation has been identified as being dual to (a variant of) the $T \bar T$ deformation of the dual CFT \cite{Giveon:2017nie}. This identifies the task of understanding a corresponding field theory deformation dual to our full solution as an interesting open question. {\itshape (See section \ref{sixvectordeformation}.)}
\item The AdS limit of our solution admits a $\tfrac12$-BPS solution of the 11-dimensional Killing spinor equation. {\itshape (See section \ref{susy}.)}
\item Finally, our solution can be used to generate new type IIA solutions by dimensional reduction (and hence other type II solutions by standard dualities). {\itshape (See section \ref{reductions}.)}
\end{itemize}

In section \ref{gentu}, we review the notions of generalised T- and U-duality that we are exploring in this paper.
We then specialise to our example involving the Euclidean 3-algebra and in section \ref{solutionderive} explain the derivation of our new solution.
We then analyse this solution (in the extremal limit) in section \ref{analysis}, explaining the points listed above.
Finally we conclude with some discussion in section \ref{discussion}. 
Appendix \ref{ingredients} lists some technical ingredients used in the main part of the paper, and appendix \ref{charges} discusses in more detail the charges of our new solution.

\section{Generalised T- and U-duality} 
\label{gentu}

\subsection{Duality and generalised geometry}

We study notions of generalised duality which can be cleanly expressed using techniques from generalised geometry and double/exceptional field theory.
Here we give a brief description of the necessary methods.
For the $d$-dimensional `internal space' $X_1$ we work with the generalised tangent bundle $T X_1 \oplus \Lambda^{(p)} T^* X_1$. Sections of this are known as generalised vectors and consist of a pair $V=(v,\omega)$ of a vector $v$ and $p$-form $\omega$.
We only need the cases $p=1$, corresponding to $O(d,d)$ generalised geometry relevant for discussing generalised T-duality in type II supergravity, and $p=2$, allowing us to describe the $\Gfour$ exceptional generalised geometry relevant for discussion of 11-dimensional supergravity when $X_1$ is four-dimensional.
In both these cases, there is a common formula for the generalised Lie derivative of generalised vectors:
\be
\mathcal{L}_V V' = ( L_v v', L_v \omega' - \iota_{v'} d \omega )\,.
\label{genlie} 
\ee
This captures the local symmetries of $X_1$, namely diffeomorphisms and gauge transformations of a $(p+1)$-form.
The geometry in the guise of the metric and this $(p+1)$-form is encoded in a generalised metric, denoted $\gM_{MN}$.
This can be factorised in terms of a generalised vielbein, $\gM_{MN} = E_M{}^A \Delta_{AB} E_N{}^B$.
If we are just interested in describing the geometry of $X_1$ then we may take $\Delta_{AB} = \delta_{AB}$, but in particular solutions on $M \times X_1$ then $\Delta_{AB}$ may depend on the coordinates of $M$ and describe scalar fields in the lower dimensional theory on $M$ obtained by reducing on $X_1$.
The inverse generalised vielbein gives a generalised frame $E_A$, providing a basis for generalised vectors.
This frame will generate an algebra under generalised Lie derivatives:
\be
\mathcal{L}_{E_A} E_B = - F_{AB}{}^C E_C\,.
\label{genpar}
\ee
If $F_{AB}{}^C$ are constant, then $E_A$ provides a \emph{generalised parallelisation}, which allows for a consistent truncation to a lower-dimensional supergravity.

A second (dual) consistent truncation then corresponds to the existence of an alternative generalised parallelisation built using a frame $\tilde E_A$ describing the generalised geometry on $X_2$.
This frame should obey the \emph{same} algebra \eqref{genpar} (possibly up to some change of basis corresponding to a constant $O(d,d)$ or $E_d$ rotation on the indices $A$).
This allows one to translate the problem of finding inequivalent dual consistent truncations to the problem of finding algebras admitting multiple solutions to the differential equations encoded in \eqref{genpar}.
As we will review below, in known variants of generalised or Poisson-Lie T- and U-duality, this can be done algorithmically within certain classes of algebras.

\subsection{Non-abelian T-duality}

The prototypical example of a generalised duality is non-abelian T-duality \cite{delaOssa:1992vci}.
This applies to spacetimes with non-abelian isometries. 
A simple example is to consider a spacetime with an $\text{S}^3$ factor (equipped with the round metric), regarded as the group manifold $\text{SU}(2)$. 
Starting with the worldsheet sigma model, we can gauge the (left) action of the group on itself and (assuming no other fields are turned on) arrive at the following dual background:
\be
ds^2 = \frac{\delta_{ij}+x_i x_j}{1+x^k x_k} \dd x^i \dd x^j \,,\quad 
B_{ij} = \frac{\epsilon_{ijk} x^k}{1+x^m x_m}\,,\quad
e^{-2\varphi} = 1+x^kx_k \,.
\label{natdgeo}
\ee
The new dual coordinates $x^i$, $i=1,2,3$ originally appear in the dualisation procedure as Lagrange multipliers imposing the flatness of the gauge field gauging the non-abelian isometry. 
Unlike in abelian T-duality, path integral arguments do not constrain the periodicity or range of these coordinates \cite{Giveon:1993ai}: we will discuss two different methods to specify the global completion of NATD solutions below.

Underlying this duality is a pair of generalised frames for the $O(d,d)$ generalised geometry.
(We describe this now with reference to the specific $\text{SU}(2)$ example, with $d=3$, but the essential features apply to $d$-dimensional group manifolds and their duals.)
The first describes the consistent truncation on the $\text{S}^3 \cong \mathrm{SU}(2)$ group manifold. 
It makes use of the following geometric data: the left-invariant forms $l^a$ and dual vectors $v_a$ obeying
\be
dl^a = \tfrac12 f_{bc}{}^a l^b \wedge l^c \,,\quad 
L_{v_a} v_b = - f_{ab}{}^c v_c \,,
\ee
where for $\mathrm{SU}(2)$ the algebra index is three-dimensional, $a=1,2,3$, and the structure constants are $f_{ab}{}^c = \epsilon_{ab}{}^c$. 
The generalised frame $E_A = \{ E_a, E^a \}$ gives a basis for sections of $T (\text{S}^3) \oplus T^* (\text{S}^3)$ with 
\be
E_a = (v_a, 0) \,,\quad 
E^a = (0,l^a) \,.
\ee
Under generalised Lie derivatives, we have the algebra \eqref{genpar} with 
\be
F_{AB}{}^C \rightarrow \{ F_{ab}{}^c = f_{ab}{}^c \,, F^{ab}{}_c = 0 \,, F_{abc} = F^{abc} = 0 \} \,.
\ee
The second generalised frame describes the dual consistent truncation on the NATD geometry \eqref{natdgeo}.
This is not a group manifold, but it can be described in terms of an underlying Poisson-Lie group structure associated to the group $U(1)^3$ (or $\mathbb{R}^3)$ with a non-trivial Poisson-Lie bivector, $\pi^{ab}$.
The latter here obeys $d \pi^{ab} = - \tilde f^{ab}{}_c \tilde l^c$, where $\tilde l^a$ are trivial left-invariant one-forms, $\tilde l^a{}_i = \delta^c_i$ (with dual vectors $\tilde v_a{}^i = \delta_a^i$) and $\tilde f^{ab}{}_c$ are dual structure constants. For the NATD of $\mathrm{SU}(2)$, these also describe the $\mathfrak{su}(2)$ Lie algebra with $\tilde f^{ab}{}_c =\epsilon^{ab}{}_c$.
We can therefore take a bivector linear in the coordinates $\pi^{ab} = - \epsilon^{ab}{}_c x^c$.
The generalised frame $\tilde E_A = \{ \tilde E_a, \tilde E^a \}$ gives a basis for sections of the extended tangent bundle of the dual geometry, with 
\be
E_a = (\tilde v_a, 0) \,,\quad 
E^a = (\pi^{ab} \tilde v_b,\tilde l^a) \,.
\label{natdframe}
\ee
Under generalised Lie derivatives, we have the algebra \eqref{genpar} with 
\be
F_{AB}{}^C \rightarrow \{ F_{ab}{}^c = 0 \,, F^{ab}{}_c = \tilde f^{ab}{}_c  \,, F_{abc} = F^{abc} = 0 \} \,.
\ee 
The use of the generalised frame \eqref{natdframe} allows for a non-geometric interpretation of the global properties of the NATD geometry.
As pointed out in \cite{Fernandez-Melgarejo:2017oyu,Bugden:2019vlj}, if we take the coordinates $x^i$ to be periodic, then under $x^i \sim x^i + \text{constant}$ the bivector $\pi^{ab}$ shifts by a constant.
Such a bivector shift can be viewed as a non-geometric $O(3,3)$ transformation.
If we patch the dual solution by such a transformation, it must be regarded as a T-fold.

It is however more common to construct global completions of NATD solutions by leveraging information about brane charges and -- for cases where there is an AdS factor in the full spacetime -- holographic duals.
To illustrate how this works, consider the example of the IIB D1-D5 near horizon solution, for which the spacetime is $\text{AdS}_3 \times \text{T}^4 \times \text{S}^3$, supported by RR flux.
The NATD dual geometry is a solution of massive IIA supergravity, with:
\be
\dd s^2 = \dd s^2_{\text{AdS}_3} + \dd s^2_{\text{T}^4} + \dd \varrho^2 + \tfrac{\varrho^2}{1+\varrho^2} \dd s^2_{\text{S}^2} \,,\quad
B = \tfrac{\varrho^3}{1+\varrho^2} \mathrm{Vol}_{\text{S}^2} \,,\quad
e^{-2\varphi} = 1+\varrho^2 \,,
\label{sfetsosthompson}
\ee
along with dual RR fields \cite{Sfetsos:2010uq}.
Here we have adopted spherical coordinates $x^i\rightarrow (\varrho, \theta,\phi)$. 
The issue of the non-compactness of dual coordinates is then concentrated in determining the range of $\varrho$.
This can be done by embedding the NATD solution into a global completion with a well-defined holographic dual and brane interpretation.
For the NATD of $\text{AdS}_5 \times \text{S}^5$ obtained in \cite{Sfetsos:2010uq} this method was demonstrated in \cite{Lozano:2016kum}, and has since been applied to many examples.
For the solution \eqref{sfetsosthompson}, the requisite completion is provided by the construction and analysis \cite{Lozano:2019emq,Lozano:2019jza,Lozano:2019zvg,Lozano:2019ywa} of a general class of massive IIA $\text{AdS}_3 \times \text{S}^2$ solutions with $3d$ $\mathcal{N}=(0,4)$ supersymmetry and an $\mathrm{SU}(2)$ structure.
The NSNS fields take the form:
\be
\begin{split}
ds^2 & = \tfrac{u}{\sqrt{h_4 h_8}} (ds^2_{\text{AdS}_3} + \tfrac{h_8 h_4}{4h_8h_4+u'{}^2} ds^2_{\text{S}^2} )
+ \sqrt{\tfrac{h_4}{h_8}} ds^2_{\text{T}^4} + \sqrt{\tfrac{h_4h_8}{u}} d\varrho^2\,, \\
B & =\tfrac12 ( -\varrho + \tfrac{uu'}{4h_8h_4 + u'{}^2} + 2 n \pi ) \mathrm{Vol}_{\text{S}^2} \,,
\end{split} 
\label{natdcomplete}
\ee
This solution exhibits the following general features found in global completions of NATD AdS solutions: 
The coordinate $\varrho$ takes values in a finite interval which is further divided into subintervals $\varrho \in [ \varrho_n, \varrho_{n+1}]$.
The functions determining the solution ($u$, $h_4$ and $h_8$) are linear in $\varrho$. They may however only be piecewise linear, and their slopes can jump from subinterval to subinterval.
The 2-form $B$ is modified by a large gauge transformation as one crosses each subinterval.
There is a (flat space) dual brane configuration, with some branes wrapping the $\varrho$ direction and others orthogonal and located at the endpoints of the subintervals.
This dual brane configuration allows for the identification of a dual quiver field theory.
The NATD solution \eqref{sfetsosthompson} can be regarded as giving the more general solution in the first subinterval, with $\varrho \in [0, \varrho_1]$, and $u \sim h_4 \sim h_8 \sim \varrho$.

Restricting to the case of vanishing Romans mass, the solutions of \cite{Lozano:2019emq,Lozano:2019jza,Lozano:2019zvg,Lozano:2019ywa} give ordinary IIA solutions which can be uplifted to M-theory \cite{Lozano:2020bxo}, giving a class of 11-dimensional $\text{AdS}_3$ solutions which we will re-encounter later.

\subsection{Poisson-Lie T- and Poisson-Lie U-duality} 

\paragraph{Poisson-Lie T-duality} 

Non-abelian T-duality can be viewed as a special case of Poisson-Lie T-duality \cite{Klimcik:1995dy,Klimcik:1995ux}, which applies to spacetimes which may lack isometries.
They instead admit an underlying Poisson-Lie group structure, involving a group $G$ equipped not only with left-invariant forms and vectors, but with a Poisson-Lie bivector. Altogether these data obey:
\be
dl^a=\tfrac12 f_{bc}{}^a l^b \wedge l^c \,,\quad
L_{v_a} v_b = - f_{ab}{}^c v_c \,,\quad
d \pi^{ab} = - \tilde f^{ab}{}_c l^c - 2 l^c f_{cd}{}^{[a} \pi^{b]d} \,,
\ee
involving simultaneously structure constants for both a Lie algebra $\mathfrak{g}$ and a `dual' Lie algebra $\tilde{\mathfrak{g}}$.
The corresponding spacetime geometry is very efficiently described by a generalised frame with: \cite{Hassler:2017yza,Demulder:2018lmj}
\be
E_a = ( v_a, 0) \,,\quad 
E^a = ( \pi^{ab}  v_b, l^a) 
\,,\quad
F_{AB}{}^C \rightarrow \{ F_{ab}{}^c = f_{ab}{}^c  \,, F^{ab}{}_c = \tilde f^{ab}{}_c  \,, F_{abc} = F^{abc} = 0 \} \,.
\ee 
The case of a standard non-abelian group manifold then has $f_{ab}{}^c \neq 0$, $\tilde f^{ab}{}_c=0$, while the NATD has the reverse.
The full doubled Lie algebra (with structure constants $F_{AB}{}^C$) here is known as the Drinfeld algebra. 
Introducing generators $T_A = \{ T_a, \tilde T^a \}$ obeying $[T_A, T_B] =F_{AB}{}^C$, we have
\be
[ T_a, T_b ] = f_{ab}{}^c T_c \,,\quad
[T_a, \tilde T^b ] = \tilde f^{bc}{}_a T_c- f_{ac}{}^b \tilde T^c
\,,\quad
[ \tilde T^a, \tilde T^b ] = \tilde f^{ab}{}_c \tilde T^c 
\ee
The algebra is further equipped with an invariant bilinear form defined by $\eta(T_a, \tilde T^b) = \delta_a^b$, and otherwise zero.
The subalgebras $\mathfrak{g} = \{T_a\}$ and $\tilde{\mathfrak{g}} = \{\tilde T^a\}$ are maximally isotropic with respect to this bilinear form, and duality at the level of the algebra involves changing one maximally isotropic subalgebra for another.
This is upgraded to a duality at the level of geometry by constructing a dual generalised frame now built using the left-invariant forms and vectors of $\tilde G = \exp \tilde{\mathfrak{g}}$  (hence the frame generates the new maximally isotropic subalgebra as its vector part), together with the corresponding Poisson-Lie bivector encoding the structure constants for $\mathfrak{g}$.

\paragraph{Poisson-Lie U-duality}

A proposal was made in \cite{Sakatani:2019zrs,Malek:2019xrf} for the algebra and generalised frames which should describe a notion of Poisson-Lie U-duality.
Let us concentrate on the case of $d=4$, for which the U-duality group is $\Gfour$.
The proposal is to consider the natural generalisation of the Poisson-Lie group to the case where the bivector is replaced by a trivector.
We then specify left-invariant forms and vectors and this trivector to obey\footnote{For simplicity, these formulae assume that $f_{ac}{}^c =0$ and that an additional scalar present in the generalised frame is constant, as is the case for the example we will study. See appendix \ref{edaframe} for more general formulae.}
\be
d l^a = \tfrac{1}{2} f_{bc}{}^a l^b \wedge l^c \,,\quad L_{v_a} v_b=-f_{ab}{}^c v_c \,, \quad
d \trivector^{abc} = \tilde f^{abc}{}_d l^d + 3 f_{ed}{}^{[a} \trivector^{bc]d} l^e \,,
\label{dPLU}
\ee
where now $a,b=1,\dots,4$.
This introduces dual structure constants $\tilde f^{abc}{}_d$ which can be viewed as defining an antisymmetric three-bracket, associated to a 3-algebra rather than an ordinary Lie algebra.

These can be used to construct a generalised frame for $\Gfour$ generalised geometry.
A generalised vector in this case is a pair of a vector and a two-form, and lies in the ten-dimensional (antisymmetric) representation of $\Gfour$.
We pick a generalised frame $E_A = (E_a, E^{ab})$, where $E^{ab} = - E^{ba}$, given by
\be
E_a = (v_a,0)\,,\quad E^{ab} = ( \trivector^{abc} v_c , l^a \wedge l^b ) \,.
\label{PLUframe}
\ee
Computing the algebra of generalised Lie derivatives \eqref{genpar} one finds an algebra dubbed the exceptional Drinfeld algebra (EDA). 
In terms of generators $T_A = (T_a, \tilde T^{ab})$, this algebra is
\be
\begin{aligned}
[T_a, T_b ] & =  f_{ab}{}^c T_c \,, \qquad
&[\tilde T^{ab}, \tilde T^{cd} ] & =   2 \tilde f^{ab[c}{}_e \tilde T^{d]e} \,,  \\
[T_a, \tilde T^{bc} ] &  =    2  f_{ad}{}^{[b} \tilde T^{c]d} - \tilde f^{bcd}{}_a T_d\,, 
& [\tilde T^{bc} , T_a ]  &=     3 f_{[de}{}^{[b} \delta_{a]}^{c]} \tilde T^{de} 
  + \tilde  f^{bcd}{}_a T_d \,.
\label{EDAcomms}
\end{aligned}
\ee
Note that these brackets are generically not antisymmetric: the EDA is generically an example of Leibniz rather than a Lie algebra.
Closure of the algebra imposes the Jacobi condition for the Lie algebra with structure constants $f_{ab}{}^c$, a cocycle condition involving both $f_{ab}{}^c$ and $\tilde f^{abc}{}_d$, and the fundamental identity for three-algebras involving just $\tilde f^{abc}{}_d$.

A notion of isotropic subalgebra exists, using now not a bilinear form but a bilinear map $\eta: \mathbf{10} \otimes_{\text{sym}} \mathbf{10} \rightarrow \mathbf{\bar{5}}$.
The subalgebra $\mathfrak{g} = \{ T_a \}$ is isotropic with respect to this definition.
However, unlike in the case of the Drinfeld double, we are not guaranteed the existence of a second, dual maximal isotropic subalgebra.
Note as well that the `symmetry' between $f$ and $\tilde f$ is now broken, and there are now more dual generators $\tilde T^{ab}$ than physical ones $T_a$.

One could nonetheless proceed to interrogate the notion of non-abelian U-duality, by starting with solutions defined by $f_{ab}{}^c \neq 0$, $\tilde f^{abc}{}_d = 0$, and dualising these, as for instance in \cite{Musaev:2020nrt}.
However, an alternative goal is to inverse the usual order, and instead look at solutions with $f_{ab}{}^c = 0$, $\tilde f^{abc}{}_d \neq 0$. 

\subsection{Dual three-algebras and beyond Poisson-Lie U-duality} 

The logic of focusing on solutions with $f_{ab}{}^c = 0$, $\tilde f^{abc}{}_d \neq 0$ is that they should be in some sense similar to the solutions generated by NATD. 
Our goal is therefore to construct examples of such solutions, verify whether they are actually `dual' to known solutions, and verify to what extent this really resembles NATD. 
Furthermore, such solutions will encode three-algebra structure constants and so are perhaps intrinsically interesting as examples of a relationship between geometry and a non-standard algebraic structure.

In \cite{Blair:2020ndg}, examples of this kind were studied, and a first look at the corresponding `3-algebra geometries' was taken, but without constructing full supergravity solutions.
A particularly interesting example is to take:
\be
\tilde f^{abc}{}_d \propto \epsilon^{abc}{}_d \,.
\label{Eucl3alg}
\ee
This is the unique Euclidean 3-algebra. 
It can be viewed as the direct generalisation of the NATD of $\mathrm{SU}(2)$, for which we had $\tilde f^{ab}{}_c = \epsilon^{ab}{}_c$.
The conditions \eqref{dPLU} can be solved by taking $l^a{}_i = \delta^a_i$, $v_a{}^i = \delta_a^i$ and a linear trivector, $\trivector^{abc} \propto \epsilon^{abc}{}_d x^d$, introducing coordinates $x^i$, $i=1,\dots,4$.
The EDA \eqref{EDAcomms} in this case turns out to be the Lie algebra $\mathrm{CSO}(4,0,1)$.

However, it turned out that it is not possible to find valid dual isotropic subalgebras within this EDA \cite{Blair:2020ndg}.
This precludes using the Poisson-Lie U-duality framework of \cite{Sakatani:2019zrs,Malek:2019xrf} to construct a dual configuration.
As noted in \cite{Blair:2020ndg}, this suggests simply that this framework may just be more restrictive than the T-duality case.
In particular, we could relax the condition that the dual isotropic be a subalgebra. 
For example, we could allow ourselves to consider alternative bases (for the same overall algebra) but for which the selected physical generators $T_a$ obey 
\be
[T_a,T_b] = \tfrac12 F_{abcd} \tilde T^{cd} \,.
\label{quasiT}
\ee 
This would be the starting point for defining a ``quasi'-EDA.\footnote{In the case of T-duality, it is possible to relax the condition that the Drinfeld double has two isotropic subalgebras, allowing to describe models with H-flux, such as those studied in the context of certain integrable deformations in \cite{Klimcik:2015gba}.}

Equivalently, we may forget about specific algebraic interpretations.
The EDA construction allows us to construct a generalised frame realising a consistent truncation from 11-dimensional SUGRA to 7-dimensional $\mathrm{CSO}(4,0,1)$ gauged SUGRA.
This consistent truncation is on a non-trivial background geometry, resulting from the generalised frame with the trivector.
However, it is already known that this gauged SUGRA can be obtained using a consistent truncation of type IIA on an $\text{S}^3$ with NSNS flux \cite{Cvetic:2000ah}.
Viewing this as M-theory on $\text{S}^3 \times I$, we have constant four-form flux, in line with the commutation relation \eqref{quasiT}.\footnote{This algebra would be explicitly realised by generalised geometric constructions of this consistent truncation \cite{Lee:2014mla, Hohm:2014qga} -- see \cite{Blair:2020ndg} for a comparison with the generalised frames of \cite{Hohm:2014qga} in particular.}
Hence, we can alternatively find `generalised U-dual' solutions by starting with solutions of type IIA supergravity to which this consistent truncation can be applied, reducing these to 7 dimensions, and then uplifting them using our EDA generalised frame for this gauging.
We will now adopt this procedure and show what it leads to for a simple brane intersecting solution.

\section{11-dimensional solution from exceptional Drinfeld algebra uplift} 
\label{solutionderive}

\subsection{Type IIA pp-F1-NS5 and reduction to 7 dimensions} 
\label{originalsoln}

We begin our solution generating procedure by taking as our original solution the non-extremal pp-F1-NS5 solution of type IIA supergravity.
After taking the five-brane decoupling limit (as reviewed in appendix \ref{ppF1NS5}) to go to the near horizon limit of the five-branes, this solution becomes:
\be
\begin{split}
ds_s^2 & = f_1^{-1} ( - f_n^{-1} W \dd t^2 + f_n ( \dd z + \tfrac{1}{2} \tfrac{r_0^2 \sinh 2 \alpha_n}{f_n r^2} \dd t )^2 ) + R^2 W^{-1} \frac{\dd r^2}{r^2} +  R^2  \dd s^2_{\text{S}^3} + \dd s^2_{\text{T}^4} \,,\\
H_{(3)} & = r_0^2 \sinh 2 \alpha_1 \frac{1}{r^3 f_1^2}  \dd t \wedge \dd z \wedge \dd r + 2 R^2 \text{Vol}_{\text{S}^3}\,,\quad
e^{-2\varphi} 
 = \tfrac{r^2}{R^2}  f_1 \,,
\end{split} 
\label{ppF1NS5limitrfinal}
\ee
where $W = 1 - \tfrac{r_0^2}{r^2}$, $R^2 \equiv N_5 l_s^2$ and
\be
f_1 = 1 + \tfrac{r_0^2 \sinh^2 \alpha_1}{r^2} \,,\quad
f_n = 1 + \tfrac{r_0^2 \sinh^2 \alpha_n}{r^2} \,,\quad
\sinh 2 \alpha_1 = \tfrac{2 N_1 l_s^2}{v} \tfrac{1}{r_0^2} \,,\quad
\sinh 2\alpha_n =  \tfrac{2 N_n l_s^4}{R_x^2 v} \tfrac{1}{r_0^2}
\,.
\label{ppF1NS5limitrfinalcharges}
\ee
Here $N_1$ is the number of F1s, $N_5$ the number of NS5s, $N_n$ the number of units of pp-wave momentum, and the four-dimensional transverse space is taken to be a torus of volume $(2\pi l_s)^4 v$.

We will be particularly interested in the extremal limit. Turning off the pp-wave contribution ($N_n=0$) the solution in this limit is 
\be
\begin{split}
ds_s^2 & = f_1^{-1} ( -  \dd t^2 + \dd z^2 )+ R^2  \frac{\dd r^2}{r^2} +  R^2  \dd s^2_{\text{S}^3} + \dd s^2_{\text{T}^4} \,,\\
H_{(3)} & = \frac{2 r_1^2 }{r^3 f_1^2}  \dd t \wedge \dd z \wedge \dd r + 2 R^2 \text{Vol}_{\text{S}^3} \,,\quad e^{-2\varphi}  = \tfrac{r^2}{R^2}  f_1 \,,
\end{split} 
\label{extremalF1NS5}
\ee
with $f_1 = 1 + \frac{r_1^2}{r^2}$, $r_1^2 = N_1 l_s^2 / v$.
This exhibits an interpolation from the near horizon region of the F1 to an asymptotic linear dilaton background.
The former corresponds to taking $f_1 = \frac{r_1^2}{r^2}$ and the solution has the form
\be
\begin{split}
ds_s^2 & = \frac{r^2}{r_1^2} ( -  \dd t^2 + \dd z^2 )+ R^2  \frac{\dd r^2}{r^2} +  R^2  \dd s^2_{\text{S}^3} + \dd s^2_{\text{T}^4} \,,\\
H_{(3)} & = \frac{2 r}{r_1^2}  \dd t \wedge \dd z \wedge \dd r + 2 R^2 \text{Vol}_{\text{S}^3} \,, \quad e^{-2\varphi}  = \tfrac{r_1^2}{R^2}  \,,
\end{split} 
\label{originalADS}
\ee
with the metric being AdS$_3\times \text{T}^4 \times \text{S}^3$.
Asymptotically, setting $f_1 =1$ and defining a coordinate $U$ by $r= R e^{U/R}$ the solution approaches the pure NS5 near horizon solution:  
\be
\begin{split}
ds_s^2 & =  -  \dd t^2 + \dd z^2 + dU^2  +  R^2  \dd s^2_{\text{S}^3} + \dd s^2_{\text{T}^4} \,,\quad
H_{(3)}  = 
 2 R^2 \text{Vol}_{\text{S}^3} \,,\quad
e^{-2\varphi}  = e^{2 U / R}  \,,
\end{split} 
\label{originalLinDil}
\ee 
with a flat metric and a linear dilaton.
We will discuss later how this interpolating behaviour is inherited by our new 11-dimensional solution.

Owing to the presence of the $\text{S}^3$ factor with accompanying NSNS flux, the background \eqref{ppF1NS5limitrfinal} can be reduced to a solution of seven-dimensional $\mathrm{CSO}(4,0,1)$ gauged maximal supergravity using the ansatz of \cite{Cvetic:2000ah}. 
The necessary part of the truncation ansatz that we need is summarised in appendix \ref{IIAS3ansatz}.
Applying this to the solution \eqref{ppF1NS5limitrfinal} gives the seven-dimensional metric, scalars $M_{ab}$ and $\Phi$, and a three-form field strength $\tilde F_{(3)}$:
\be
\begin{split}
ds_7^2& = (r/R)^{4/5} f_1^{2/5} \left( f_1^{-1}(- f_n^{-1} W \dd t^2 + f_n ( \dd z + \tfrac{1}{2} \tfrac{r_0^2 \sinh 2 \alpha_n}{f_n r^2} \dd t )^2 ) + R^2  W^{-1} \frac{\dd r^2}{r^2} + \dd s^2_{\text{T}^4} \right)\,,\\
M_{ab} & = \delta_{ab} \,,\quad \Phi  = f_1^{-4/5} (r/R)^{-8/5} \,,\quad
\tilde F_{(3)} = r_0^2 \sinh 2 \alpha_1\frac{1}{f_1^2 r^3} \dd t \wedge \dd z  \wedge \dd r \,.
\end{split}
\label{sevendata}
\ee
All other fields in the ansatz are vanishing.
We next identify the data of \eqref{sevendata} with the appropriate $\Gfour$ covariant fields of the $\mathrm{CSO}(4,0,1)$ gauged supergravity.
Take $\fA = (a,5)$ to be a five-dimensional fundamental $\Gfour$ index, and let $A$ denote a ten-dimensional index for the antisymmetric representation. 
The $\Gfour$ covariant fields are: the $\Gfour$-invariant metric $ds^2_7$, a scalar matrix $\gM_{\fA \fB}$ parametrising the coset $\Gfour/\Hfour$, and gauge fields in $\Gfour$ representations.
The latter include a one-form $\Aa_\mu{}^A$, in the 10-dimensional representation and a two-form $\Ab_{\mu\nu \fA}$ in the five-dimensional representation, with corresponding field strengths $\Fa_{\mu\nu}{}^A$ and $\Fb_{\mu\nu\rho \fA}$.
The fields \eqref{sevendata} provide a non-trivial scalar matrix and three-form field strength:
\be
\mathcal{M}_{\fA \fB} = \begin{pmatrix}
\Phi^{-\tfrac{1}{4}} \delta_{ab} & 0 \\
0 & \Phi
\end{pmatrix} \,,
\quad
\Fb_{(3) \fA} = ( 0 ,\tilde F_{(3)}) \,.
\ee

\subsection{11-dimensional uplift via exceptional field theory} 
\label{exftuplift}

Having mapped our solution to seven-dimensional gauged supergravity, we now uplift it to a \emph{different} higher-dimensional solution using a distinct consistent truncation corresponding to the exceptional Drinfeld algebra realisation of the $\mathrm{CSO}(4,0,1)$ algebra \cite{Blair:2020ndg}.
This makes use of the $\Gfour$ covariant reformulation of supergravity provided by $\Gfour$ exceptional field theory (ExFT). 
To describe this uplift, let $y^\mu$ denote seven-dimensional coordinates describing the solution \eqref{sevendata}.
We introduce an $\Gfour$-valued \emph{generalised frame field} denoted by $\tilde E^M{}_A(x)$ in the ten-dimensional representation or by $\tilde E^{\fM}{}_{\fA}(x)$ in the five-dimensional representation, as well as a scalar function $\Delta(x)$.
These depend on a set of four-dimensional coordinates $x^i$, $i=1,\dots,4$, which will describe the internal space of the new eleven-dimensional solution. 
The new eleven-dimensional solution has a simple $\Gfour$ covariant construction: we define the ExFT external metric, generalised metric and field strengths by
\be
\begin{split}
g_{\mu\nu}(y,x) &= \Delta^2(x) g_{\mu\nu}(y) \,,\quad
\gM_{\fM\fN}(y,x) = \tilde E^{\fA}{}_{\fM}(x) \tilde E^{\fB}{}_{\fN}(x) \gM_{AB}(y) \,,\\
\Fa_{(2)}{}^{M}(y,x) &= \Delta (x) \tilde E^M{}_A(x) \Fa_{(2)}{}^{A}(y) \,,\quad \Fb_{(3) \fM}(y,x)  = \Delta^2 (x) \tilde E^{\fA}{}_{\fM}(x) \Fb_{(3) \fA}(y) \,.
\end{split}
\label{ScherkSchwarz}
\ee
It is in fact the combination $E^M{}_A \equiv \Delta \tilde E^M{}_A$ that must be used to construct the generalised frame \eqref{PLUframe} obeying the generalised parallelisation condition \eqref{genpar}.
To realise the $\mathrm{CSO}(4,0,1)$ algebra we take trivial left-invariant forms and vectors, $l^a{}_i =\delta^a_i$, $v_a{}^i = \delta_a^i$, and a trivector linear in the coordinates $x^i$.
The choice $\trivector^{abc} = \tfrac{1}{R} \epsilon^{abc}{}_d x^d$ reproduces the $\mathrm{CSO}(4,0,1)$ algebra and the scalar potential arising from the truncation of type IIA on an S$^3$ of radius $R$ (see appendix \ref{edaframe}). Note here we can use $\delta^a_i$ to identify curved and flat indices here, for convenience.
In terms of the five-dimensional representation of $\Gfour$, this gives a generalised frame:
\be
\tilde E^{\fA}{}_{\fM} =\begin{pmatrix}
\delta^{a}_{m} & 0 \\
- \tfrac{x_{m}}{ R} & 1
\end{pmatrix}\,,\quad
\Delta = 1 \,.
\label{CSOframe}
\ee
Using \eqref{CSOframe} and \eqref{ScherkSchwarz} applied to the background arising from the pp-F1-NS5 solution, we obtain a generalised metric and three-form of the form
\begin{equation}
\mathcal{M}_{\fM \fN}
=\begin{pmatrix}
\Phi^{-\tfrac{1}{4}} \delta_{mn}+\Phi \tfrac{1}{R^2 } x_{m}x_{n} & -\Phi  \tfrac{1}{R } x_{m} \\
-\Phi\tfrac{1}{R } x_{n} & \Phi
\end{pmatrix}\,,\quad
\Fb_{(3) \fM} = ( - \tfrac{x_m}{R } \tilde F_{(3)} , \tilde F_{(3)}  )\,,
\end{equation}
while the seven-dimensional ExFT external metric is unchanged.
It is then a straightforward matter to convert this to a standard description in terms of the eleven-dimensional metric and four-form field strength using the known ExFT dictionary (see for instance the review \cite{Berman:2020tqn}), summarised in appendix \ref{exft}.

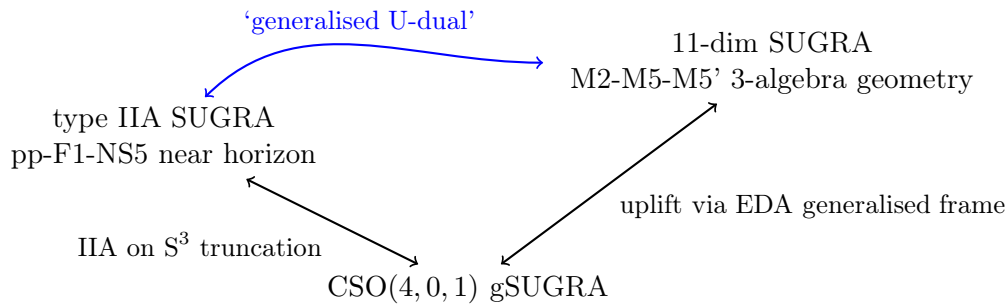
\begin{figure}[h]
\centering
\begin{tikzpicture}

\draw (-4,2) node (A)[text width=6cm,align=center] {type IIA SUGRA \\pp-F1-NS5 near horizon};

\draw (0,0) node (seven) {$\mathrm{CSO}(4,0,1)$ gSUGRA};

\draw (4,3) node (M) [text width=5.75cm,align=center] {11-dim SUGRA \\ M2-M5-M5' 3-algebra geometry};

\draw[thick,<->] (A) to node [midway, below left] {\small IIA on $\text{S}^3$ truncation} (seven);
\draw[thick,<->] (seven) to node [midway, below right] {\small uplift via EDA generalised frame} (M);

\draw[thick,<->,blue] (A) to[out=45,in=180] node [midway, above] {\small `generalised U-dual'} (M);

\end{tikzpicture}
\label{map}
\caption{The relationship between our solutions}
\end{figure}

\subsection{Resulting solution}

Using equation \eqref{SL5genm} for the parametrisation of the generalised metric allows one to obtain the new internal four-dimensional metric and three-form, with the latter given by
\be
C_{ijk} = - \frac{ \epsilon_{ijkl} R  x^l}{ r^2 f_1  + x_m x^m } \,.
\label{upliftC}
\ee
As there is no ExFT one-form present, the Kaluza-Klein vector $A_\mu{}^i$ vanishes, and using \eqref{11Decomp} one obtains the full 11-dimensional metric
\be
\begin{split} 
\dd s_{11}^2  &= (r^2 f_1 +  x_k x^k)^{1/3} \Bigg[
\frac{(r^2 f_1)^{1/3}}{R^{4/3}}\Big(  
ds^2_{M_3}
 + \dd s^2_{\text{T}^4} \Big)
 +  R^{2/3} (r^2f_1)^{1/3} \frac{ \left(\delta_{ij} + \tfrac{ x_ix_j}{  r^2 f_1} \right) }{r^2 f_1 + x^k x_k }\dd x^i \dd x^j\Bigg]
\end{split}
\label{Uplifted1}
\ee
where
\be
ds^2_{M_3} = f_1^{-1} ( - f_n^{-1} W \dd t^2 + f_n ( \dd z + \tfrac{1}{2} \tfrac{r_0^2 \sinh 2 \alpha_n}{f_n r^2} \dd t )^2 ) + \tfrac{R^2 \dd r^2}{r^2 W}\,.
\ee
The three-form \eqref{upliftC} and the new four-dimensional part of the metric in equation \eqref{Uplifted1} closely resemble the two-form and metric appearing in the NATD of $\text{S}^3$ \eqref{natdgeo}, but now in one dimension higher (this is easiest to see by setting $r^2f_1=1$).

To complete the solution, we use \eqref{deftildeF} to extract the remaining components of the four-form field strength (via a dualisation, as $\cH_{\mu\nu\rho 5}$ directly gives components of the seven-form field strength).
This gives a total four-form field strength:
\be
\begin{split}
F_{(4)} & = 
\frac{r_0^2 \sinh 2 \alpha_1}{ (r^2 f_1)^2} \frac{r x_i}{R} \dd t \wedge \dd z \wedge \dd r \wedge \dd x^i 
\rpm \frac{r_0^2 \sinh 2\alpha_1}{R^3}\mathrm{Vol}_{\text{T}^4} 
\\ & \qquad
+ \frac{R \tfrac{1}{4!}\epsilon_{ijkl} \dd x^i \wedge \dd x^j \wedge \dd x^k }{(r^2 f_1 + x_px^p)^2} \wedge \left(
(4 r^2 f_1 + 2 x_q x^q)  \dd x^l 
- 4 x^l \partial_r (r^2 f_1) \dd r  
\right)\,.
\end{split} 
\label{Uplifted2}
\ee 
The dual seven-form field strength is\footnote{We define the Hodge dual of a $p$-form $F$ via
\be
( \star F)_{\mu_1 \dots \mu_{D-p} }  
= \frac{1}{p!} \sqrt{|g|}  \epsilon_{\mu_1 \dots \mu_{D-p} \nu_1 \dots \nu_p} g^{\nu_1\rho_1} \dots g^{\nu_p \rho_p} F_{\rho_1 \dots \rho_p}
\,,
\ee
where $\epsilon_{\mu_1\dots \mu_D}$ denotes the Levi-Civita symbol $\epsilon_{01\dots D-1}= +1$.
This obeys $\star \star F = (-1) (-1)^{p(D-p)} F$.
}
\be
\begin{split}
\star F_{(4)} & = 
\frac{r_0^2 \sinh 2 \alpha_1}{r^2 f_1 +  x_p x^p } \frac{ \epsilon_{ijkl} x^l}{ R^2} \tfrac{1}{3!} \dd x^i \wedge \dd x^j \wedge \dd x^k \wedge \mathrm{Vol}_{\text{T}^4}\\
& \quad \rpm \frac{r_0^2 \sinh 2\alpha_1}{r f_1 ( r^2 f + x_p x^p )} \tfrac{1}{4!} \epsilon_{ijkl} \dd t \wedge \dd z \wedge \dd r \wedge \dd x^i \wedge \dd x^j \wedge \dd x^k \wedge \dd x^l 
\\ & \quad + \frac{2r}{R^4 } ( 2 r^2 f_1 + x_k x^k ) \dd t \wedge \dd z \wedge \dd r \wedge \mathrm{Vol}_{\text{T}^4}
 + \frac{r^2 W}{R^3 r f_1} \frac{x_i}{R } \partial_r(r^2 f_1) \dd t \wedge \dd z \wedge \dd x^i  \wedge \mathrm{Vol}_{\text{T}^4}\,.
\end{split} 
\ee 
Note that $(\star F_{(4)})_{ijk y^1 \dots y^4} = \rmp C_{ijk} F_{y^1 \dots y^4}$.
We have $\dd \star F_{(4)} = \rmp \tfrac{1}{2} F_{(4)} \wedge F_{(4)}$.

\section{Analysis of the extremal 11-dimensional solution}
\label{analysis}

We now restrict to the extremal limit and set the pp-wave contribution to zero, making the replacements $W \rightarrow 1$, $f_1 \rightarrow 1+\tfrac{r_1^2}{r^2}$, $r_0^2 \sinh 2\alpha \rightarrow 2 r_1^2$, $\alpha_n \rightarrow 0$.
We can also simplify the form of our solution by appropriately rescaling the coordinates as well as the metric and three-form so as to effectively set the constants $r_1$ and $R$ equal to 1.\footnote{To be precise: this involves setting $(t,z,y^I) = R(\tilde t, \tilde z, \tilde y^I)$ and $(r, x^i) = r_1(\tilde r, \tilde x^i)$, such that $\dd s_{11}^2 = R^{2/3} r_1^{4/3} \widetilde{\dd s}_{11}^2$, $F_{(4)} = R r_1^2 \widetilde{F}_{(4)}$. We then work with  $\widetilde{\dd s}_{11}^2$ and $\widetilde{F}_{(4)}$, in which no dimensionful constants appear (and drop tildes). This scaling of the metric and gauge field is a symmetry of the equations of motion (the trombone). We can also introduce this scaling directly into the ExFT frame by introducing a constant parameter $\alpha$ as in appendix \ref{edaframe}.} 

\subsection{Solution as a U-fold} 
\label{Ufold}

Having made these simplifications, we henceforth study the following solution of 11-dimensional supergravity:
\be
\begin{split} 
\dd s_{11}^2  &= (r^2 f_1 +  x_k x^k)^{1/3} (r^2 f_1)^{1/3} \Big(  
 f_1^{-1} ( -  \dd t^2 + \dd z^2 ) +  \frac{\dd r^2}{r^2} + \dd s^2_{\text{T}^4} \Big)
\\ & \quad
 + ( r^2 f_1 +  x_k x^k)^{-2/3}  (r^2f_1)^{1/3} \left(\delta_{ij} + \frac{ x_ix_j}{  r^2 f_1} \right) \dd x^i \dd x^j
\\
F_{(4)} & = 
\frac{2rx_i}{ (r^2 f_1)^2}  \dd t \wedge \dd z \wedge \dd r \wedge \dd x^i 
\rpm 2 \mathrm{Vol}_{\text{T}^4} 
+ \frac{ ( 4 r^2 f_1 + 2 x_q x^q ) }{ (r^2 f_1 + x_p x^p )^2}\tfrac{1}{4!} \epsilon_{ijkl} \dd x^i \wedge \dd x^j \wedge \dd x^k \wedge \dd x^l 
\\ & \qquad + \frac{x^l \partial_r( r^2 f_1)}{ (r^2 f_1 +  x_px^p)^2}  \tfrac{1}{3!} \epsilon_{ijkl} \dd r \wedge \dd x^i \wedge \dd x^j \wedge \dd x^k \,.
\end{split} 
\label{UpliftedSimpler}
\ee 
with $f_1 = 1 + \tfrac{1}{r^2}$.

If we take the $x^i$ coordinates to be periodic, this should be identified as a U-fold.
This is analogous to the interpretation of NATD solutions as T-folds suggested in \cite{Fernandez-Melgarejo:2017oyu,Bugden:2019vlj}.
For our solution, this U-fold interpretation follows from the form of the EDA frame, which features a trivector depending linearly on the coordinates $x^i$.
The patching for $x^i \sim x^i +$ constant amounts therefore to a shift of this trivector, which is a non-trivial non-geometric U-duality transformation.
From \eqref{CSOframe} we have
\be
\tilde E^{\fA}{}_{\fM} ( x^i +R n^i) = \tilde E^{\fA}{}_{\fN} U^{\fN}{}_{\fM} \,,\quad
U^{\fN}{}_{\fM} = \begin{pmatrix} \delta^n_m & 0 \\ - n_m & 1 \end{pmatrix} \,.
\label{uduality} 
\ee
If $n_i = \delta_{ij} n^j$ are integers the matrix defines an $\mathrm{SL}(5;\mathbb{Z})$ U-duality transformation.
We can describe its action on the four-dimensional internal geometry with metric $\phi_{ij}$ and three-form $C_{ijk}$ using the generalised metric $\gM_{\fM\fN}$, which is a symmetric unit determinant five-by-five matrix, parametrised by the metric and three-form as in \eqref{SL5genm}.
Under $U \in \mathrm{SL}(5)$, this transforms as $\gM_{\fM\fN} \rightarrow U^{\fP}{}_{\fM} U^{\fQ}{}_{\fN} \gM_{\fP \fQ}$.
In the present case, we factorise $\gM_{\fM \fN}(y,x) = \tilde E^{\fA}{}_{\fM} (x)\gM_{AB}(y) \tilde E^{\fB}{}_{\fN}(x)$, where as above $y$ denotes 7-dimensional coordinates.
This manifestly shows that the generalised metric and hence four-dimensional metric and three-form together transform under the U-duality transformation, or monodromy, in \eqref{uduality}, for periodic $x^i$.

Associated to this U-fold interpretation is the fact that one can interpret the three-algebra structure constants as (non-geometric) M-theory Q-flux \cite{Blair:2014zba}.
This is here defined by $Q_a{}^{bcd} \sim \partial_a \trivector^{bcd} \sim \tilde f^{bcd}{}_a$.

We will not further pursue this U-fold interpretation, but now focus on ordinary geometric properties of the solution \eqref{UpliftedSimpler}.

\subsection{Solution in spherical coordinates and brane charges}
\label{solnsph}

We can rewrite the solution \eqref{UpliftedSimpler} by changing to spherical coordinates, letting $x^i = \rho \mu^i$ with $\mu^i \mu^j \delta_{ij} = 1$.
This is what is usually done for solutions obtained via non-abelian T-duality.
The possible non-compactness of the solution will now be determined by the range of $\rho$.
In these coordinates, the metric and field strength of \eqref{UpliftedSimpler} have the form
\be
\begin{split} 
\dd s_{11}^2  &= (r^2 f_1 + \rho^2)^{1/3}  (r^2 f_1)^{1/3} 
\left(  
\frac{1}{f_1} ( - \dd t^2 + \dd z^2 ) 
+    \frac{\dd r^2}{r^2}+ \dd s^2_{\text{T}^4} + \frac{\dd \rho^2}{r^2 f_1}  
\right)
\\ & \quad
 + ( r^2 f_1 +  \rho^2)^{-2/3} (r^2f_1)^{1/3} \rho^2 \dd s^2_{\text{S}^3} 
\,,
\\
F_{(4)} & = 
\frac{2 r \rho }{(r^2 f_1)^2}\dd t \wedge \dd z \wedge \dd r \wedge \dd\rho 
\rpm 2 \mathrm{Vol}_{\text{T}^4}
\\ & \qquad + 
\frac{( 4 r^2 f_1 + 2  \rho^2 ) }{(r^2 f_1 +  \rho^2 )^2}  \rho^3 \dd\rho \wedge \mathrm{Vol}_{\text{S}^3} 
- \frac{\rho^4 \partial_r ( r^2 f_1)}{(r^2 f_1 +  \rho^2)^2}   \dd r \wedge \mathrm{Vol}_{\text{S}^3}  \,.
\end{split} 
\label{solnSphericalSimpler}
\ee 
The dual field strength is
\be
\begin{split}
\star F_{(4)} & = 
-\frac{2 \rho^4}{r^2 f_1 +  \rho^2 } \mathrm{Vol}_{\text{S}^3} \wedge \mathrm{Vol}_{\text{T}^4}  
 \rpm \frac{2\rho^3}{r f_1 ( r^2 f_1 + \rho^2 )}  \dd t \wedge \dd z \wedge \dd r \wedge \dd \rho \wedge  \mathrm{Vol}_{\text{S}^3} 
\\ & \quad + 2r ( 2 r^2 f_1 +  \rho^2) \dd t \wedge \dd z \wedge \dd r \wedge\mathrm{Vol}_{\text{T}^4} 
+ \frac{r\rho}{f_1} \partial_r(r^2 f_1) \dd t \wedge \dd z \wedge \dd \rho  \wedge \mathrm{Vol}_{\text{T}^4}\,.
\end{split} 
\ee 
We can discuss the possible M2 and M5 brane charges carried by this solution.
These will be given by integrals\footnote{It is possible to make this more exact and to in particular require quantised charges: we defer this discussion to appendix \ref{charges}.}
\be
q_{M2} = \int J_{\text{Page}} \,,\quad
q_{M5} = \int F_{(4)} \,,
\ee
where 
the Page charge density for M2 branes is $J_{\text{Page}} = \star F_{(4)} \rpm \tfrac{1}{2} C_{(3)} \wedge F_{(4)}$.
Let us consider the latter.
Let $C_{\text{sphere}}$ and $C_{\text{torus}}$ denote the restriction of the three-form to the sphere and torus respectively. 
We have
\be
C_{\text{torus}} \wedge \dd C_{\text{sphere}} +
C_{\text{sphere}}\wedge \dd C_{\text{torus}}
= \dd ( C_{\text{torus}} \wedge C_{\text{sphere}} ) + 2 C_{\text{sphere}}\wedge \dd C_{\text{torus}} \,.
\label{tricky}
\ee
An explicit choice of potential is:
\be
C_{(3)} = 
\frac{\rho}{f_1} \dd t\wedge \dd z\wedge \dd\rho
\rpm 2 c_{(3)}
 + \frac{\rho^4}{r^2 f_1 + \rho^2} \mathrm{Vol}_{\text{S}^3}\,,
\label{explicitpot}
\ee
where $\dd c_{(3)} = \mathrm{Vol}_{\text{T}^4}$. For this potential, the second term in \eqref{tricky} cancels with the contribution from $\star F_{(4)}$ such that
$J_{\text{Page}} = -  
\dd \left( c_{(3)} \wedge \frac{
\rho^4}{r^2 f_1 + \rho^2} \mathrm{Vol}_{\text{S}^3}\right)$ 
and therefore is a total derivative. Hence the M2 charge vanishes up to large gauge transformations.
In particular we can consider a large gauge transformation given by 
\be
C_{(3)} \rightarrow C_{(3)} + 4\pi j  
\mathrm{Vol}_{\text{S}^3}
\label{largegauge}
\ee
such that $T_{M2} \int C_{(3)} \rightarrow T_{M2} \int C_{(3)}+ 2\pi j$, with $j \in \mathbb{Z}$.
Using \eqref{tricky} this means
\be
J_{\text{Page}} \rightarrow 8 \pi j 
\mathrm{Vol}_{\text{S}^3} \wedge \mathrm{Vol}_{\text{T}^4} \,,
\ee
which generates a non-trivial M2 charge. 

Next we consider the possible M5 brane charge.
We firstly have a non-trivial M5 charge given by integrating $F_{(4)}$ against the torus.
The M2 charge generated by the above large gauge transformation will be proportional to this M5 charge.

A further M5 charge, denoted M5', could be obtained by integrating $F_{(4)}$ over a four-cycle involving $r$, $\rho$ and the sphere directions. 
Following closely the analysis of NATD solutions in \cite{Terrisse:2018hhf}, we look for a path in the $(r,\rho)$ directions such that the three-sphere shrinks to zero size at beginning and end of the path, giving a closed four-cycle.
This happens at $\rho =0$; suppose it also happens for some value of $r=r_s$.
Then a possible integration is to integrate from $\rho=0$ to $\rho=\bar\rho$ at fixed $r=\bar r$, and then integrate at fixed $\bar \rho$ from $\bar r$ to $r=r_s$.
Letting $C(\rho,r)= \frac{\rho^4}{r^2 f_1 + \rho^2}$ we would then have
\be
\begin{split} 
\int_{\rho=0}^{\rho=\bar{\rho}} F_{(4)} \Big|_{r=\bar{r}} +
\int_{r=\bar{r}}^{r=r_s} F_{(4)} \Big|_{\rho=\bar{\rho}}
& = 2\pi^2 \left(
C(\bar{\rho}, \bar{r}) - C(0,\bar{r}) + C(\bar{\rho},r_s) - C(\bar{\rho},\bar{r}) 
\right)
\\ & = 2\pi^2 ( C(\bar{\rho}, r_s) - C(0,\bar{r}) )
= \frac{2 \pi^2 \bar{\rho}^4}{r_s^2 f_1(r_s) + \bar{\rho}^2}\,.
\end{split}
\ee
This is independent of $\bar r$.
The issue is now whether one can find a closed four-cycle with the above properties.
This issue is linked to the question of finding a global completion of the solution \eqref{solnSphericalSimpler}.
Indeed, for the full metric \eqref{solnSphericalSimpler} there is no way to close the cycle to give a non-zero value for the above integration.
This is a signal that one needs additional ingredients, such as will be discussed in the next subsection at least for the AdS limit.

For the solution with $f_1=1$, that we would obtain by starting with the pure NS5 near horizon solution \eqref{originalLinDil}, extra ingredients are not needed.
Our new 11-dimensional solution in this case has the form:
\be
\begin{split} 
\dd s_{11}^2  &= (r^2  + \rho^2)^{1/3}  r^{2/3}
\left(  
 - \dd t^2 + \dd z^2 +    \frac{\dd r^2}{r^2}+ \dd s^2_{\text{T}^4} + \frac{\dd \rho^2}{r^2}  
\right)
 + ( r^2 +  \rho^2)^{-2/3} r^{2/3}  \rho^2 \dd s^2_{\text{S}^3} 
\,,
\\
F_{(4)} & = \dd \left( \frac{\rho^4}{r^2 + \rho^2} \mathrm{Vol}_{\text{S}^3} \right)\,.
\end{split} 
\label{dualtoNS5}
\ee 
A valid choice for the above four-cycle is to take $r_s = 0$ for which 
\be
q_{M5} = 2\pi^2 \bar \rho^2 \,.
\ee
Restoring dimensionful constants and requiring this to give a quantised brane charge provides one possible way to determine the range of $\rho$, fixing it to lie in the finite interval $\rho \in [0,\bar \rho]$.

\subsection{AdS limit and holographic completion}
\label{adslimit}

The AdS limit amounts to setting $r^2 f_1 = 1$ in the solution \eqref{solnSphericalSimpler}:
\be
\begin{split} 
\dd s_{11}^2  &= (1 + \rho^2)^{1/3}  
\left(  
\dd s_{\text{AdS}_3}^2  + \dd \rho^2 + \dd s^2_{\text{T}^4}
\right)
	+ ( 1 +  \rho^2)^{-2/3} \rho^2 \dd s^2_{\text{S}^3} 
\,,
\\
F_{(4)} & = 
2 \rho \mathrm{Vol}_{\text{AdS}_3}\wedge \dd\rho 
\rpm 2 \mathrm{Vol}_{\text{T}^4}
+ \frac{( 4  + 2  \rho^2 ) }{(1+  \rho^2 )^2}  \rho^3 \dd\rho \wedge \mathrm{Vol}_{\text{S}^3} \,.
\end{split} 
\label{AdSlimitSimpler}
\ee 
In terms of the original F1-NS5 solution \eqref{extremalF1NS5}, this corresponds to going to the near horizon region also of the F1.

The solution \eqref{AdSlimitSimpler} fits into a general class of M-theory $\text{AdS}_3$ solutions constructed in \cite{Lozano:2020bxo}.
These solutions are of the form $\text{AdS}_3 \times \text{S}^3/\mathbb{Z}_k \times \text{CY}_2$ foliated over an interval.
They are closely related to the $\text{AdS}_3 \times \text{S}^2$ solutions \eqref{natdcomplete} in massive IIA which provide a way to complete the NATD of $\text{AdS}_3 \times \text{T}^4 \times \text{S}^3$.
Restricting this class of solutions to ordinary IIA (by setting $h_8$ constant) allows for an uplift to M-theory.
The resulting solutions presented in \cite{Lozano:2020bxo} read as follows:
\be
\begin{split} 
\dd s_{11}^2
 & = \Delta\left(\tfrac{u}{\sqrt{\hat{h}_4 h_8}} \dd s_{\text{AdS}_3}^2+\sqrt{\tfrac{\hat{h}_4}{h_8}} \dd s_{\text{CY}_2}^2+\tfrac{\sqrt{\hat{h}_4 h_8}}{u} \dd \varrho^2
\right)+\frac{h_8^2}{\Delta^2} \dd s^2_{\text{S}^3/\mathbb{Z}_k} \, ,\quad 
 \Delta =\tfrac{h_8^{1/2}\left(\hat{h}_4h_8+\tfrac14 u'^2\right)^{1/3}}{\hat{h}_4^{1/6}u^{1/3}}\, ,
\\
F_{(4)}&= -\dd\left(\tfrac{uu'}{2\hat{h}_4}+2\varrho h_8\right)\wedge \text{{Vol}}_{\text{AdS}_3} -\partial_{\varrho}\hat{h}_4\;{\text{Vol}}_{\text{CY}_2}  
+2h_8\;\dd\left(-\varrho+\tfrac{u u'}{4\hat{h}_4h_8+u'^2}\right)\wedge {\text{Vol}}_{\text{S}^3/\mathbb{Z}_k} \,,
 \\
\end{split}
\label{targetAdS}
\ee
where the quotiented 3-sphere is written as an S$^1$ Hopf fibration over an S$^2$
\begin{eqnarray}
\dd s^2_{\text{S}^3/\mathbb{Z}_k}=\frac{1}{4}\left[\left(\tfrac{ \dd \psi}{k}+\eta\right)^2+\dd s^2_{\text{S}^2}\right]\,,\quad \dd \eta={\text{Vol}}_{\text{S}^2}\,.
\label{saza}
\end{eqnarray}
The functions $u$ and $\hat h_4$ are again linear functions of $\varrho$, but $h_8$ is given by $h_8 = k$ an integer.

To match this to our solution \eqref{AdSlimitSimpler}, we relate our radial spherical coordinate $\rho$ to the coordinate $\varrho$ appearing in \eqref{targetAdS} via:
\be
\rho^2 = 2 \varrho \,.
\label{match1}
\ee
This allows us to write \eqref{AdSlimitSimpler} as
\be
\begin{split} 
\dd s_{11}^2  &= (1 + 2 \varrho )^{1/3}  
\left(  
\dd s_{\text{AdS}_3}^2  + \tfrac{\dd \varrho^2}{2\varrho} + \dd s^2_{\text{T}^4}
\right)
	+ ( 1 + 2 \varrho)^{-2/3} 2\varrho\dd s^2_{\text{S}^3} 
\,,
\\
F_{(4)} & = 
2  \mathrm{Vol}_{\text{AdS}_3}\wedge \dd\varrho 
\rpm 2 \mathrm{Vol}_{\text{T}^4}
+ \frac{8(1+\varrho)}{(1+  2\varrho)^2} \varrho \dd\varrho \wedge \mathrm{Vol}_{\text{S}^3} \,.
\end{split} 
\label{AdSlimitLinearSimpler}
\ee 
It is straightforward to confirm that the solution \eqref{AdSlimitLinearSimpler} is included in the class of solutions \eqref{targetAdS} for:\footnote{To match precisely, we need to take into account some freedom to change signs of components of our four-form field strength, e.g. the overall sign $C_{(3)}\rightarrow-C_{(3)}$ is a matter of convention/orientation, we may also flip the sign of a torus coordinate, or change the sign of the electric $B$-field components of the original F1-NS5 solution.}
\be
k=1\,,\quad  u(\varrho)=\hat h_4(\varrho) =2 \varrho\,,
\label{match2}
\ee 
giving $\Delta = (1+2\varrho)^{1/3} / (2\varrho)^{1/2}$, and taking the CY${}_2$ to correspond to $\text{T}^4$ specifically (we could equally well have considered our solution on either $\text{T}^4$ or K3 from the beginning).

The general class of solutions \eqref{targetAdS} then has the necessary properties needed to provide a global completion and holographic dual of the AdS limit of our solution.
As specified in \cite{Lozano:2020bxo}, one considers the following set-up.
The coordinate $\varrho$ takes values in a finite interval $\varrho \in [0, 2\pi (P+1)]$, which is divided into subintervals $\varrho \in [2\pi j, 2\pi(j+1)]$ for $j=0,\dots P$.
The function $u$ is linear in $\varrho$, while $\hat h_4$ is piecewise linear, with its slope jumping from subinterval to subinterval. 
It further is taken to obey $\hat h_4(0) = \hat h_4(2\pi(P+1)) = 0$, which has the effect of `ending' the space at the endpoints of the interval (and allows for the computation of M5' brane charge by integrating the four-form flux on the full $\rho$ interval and $\text{S}^3$).
The 3-form $C_{(3)}$ is modified by a large gauge transformation (of the form \eqref{largegauge}) as one crosses the endpoints of each subinterval.
There is a (flat space) underlying brane configuration, involving M5 branes wrapping the $(t,z,r)$ and $\text{S}^3$ directions, M5' branes wrapping the $(t,z)$ and torus directions, and positioned at $\varrho=2\pi j$, and M2 branes wrapping the $(t,z,\varrho)$ directions stretched between these M5 branes.
This dual brane configuration allows for the identification of a dual quiver field theory, described in \cite{Lozano:2020bxo}.
Our solution \eqref{AdSlimitSimpler} can be regarded as giving the more general solution only in the first subinterval, with $\varrho \in [0, 2\pi]$.
This is exactly analogous to the situation with NATD solutions, and shows that our solution based on dual three-algebra rather than Lie algebra structure constants admits a similar holographic interpretation.

\subsection{Full solution as a six-vector deformation of AdS limit} 
\label{sixvectordeformation}

We now return to the full solution \eqref{solnSphericalSimpler}, in order to explain how it can be viewed as a particular interpolation away from, or deformation of, its AdS${}_3$ limit.
To show this, it is helpful (though not strictly necessary) to introduce a dimensionless parameter $\lambda$ by rescaling the AdS coordinates as
\be
t \rightarrow \lambda^{-1/2} t \,,\quad
z \rightarrow \lambda^{-1/2} z \,,\quad
r \rightarrow \lambda^{+1/2} r \,.
\label{rescaling} 
\ee
The parameter $\lambda$ now serves as a book-keeping device for describing the deformation of the AdS limit, which corresponds to $\lambda = 0$. 
The function $f_1$ is now $f_1 = 1 + \tfrac{1}{\lambda r^2}$ and hence the $\lambda \rightarrow 0$ limit picks out the near horizon region where one drops the constant term.
Evidently for $\lambda=0$ the rescaling \eqref{rescaling} is singular, but nonetheless the metric and field strength are well-defined.
Explicitly, one has:
\be
\begin{split} 
\dd s_{11}^2  &=
 (1+\rho^2 + \lambda r^2)^{1/3} (1+\lambda r^2)^{-2/3} 
 \left( r^2 (-\dd t^2 + \dd z^2) +  d \rho^2 \right)
\\&\qquad +  (1+\rho^2 + \lambda r^2)^{1/3} (1+\lambda r^2)^{1/3} \left( \frac{\dd r^2}{r^2} + \dd s^2_{\text{T}^4} \right)
 \\& \qquad +   (1+\rho^2 + \lambda r^2)^{-2/3} (1+\lambda r^2)^{1/3} \rho^2 \dd s^2_{\text{S}^3} \,,\\
 F_{(4)} & = \frac{2r\rho}{(1+\lambda r^2)^2} \dd t \wedge \dd z \wedge \dd r \wedge \dd \rho \rpm 2 \mathrm{Vol}_{\text{T}^4} 
 + d \left(\frac{\rho^4}{1+\lambda r^2 + \rho^2} \mathrm{Vol}_{\text{S}^3} \right)\,.
\end{split} 
\label{solutionDeformation}
\ee
This indeed reduces to the AdS limit \eqref{AdSlimitSimpler} for $\lambda = 0$.
For $\lambda \neq 0$ one has the full solution (in which we can always undo the rescaling by setting $\lambda=1$).

The solution \eqref{solutionDeformation} with finite $\lambda$ can be expressed as an $\Gsix$-valued deformation of the $\lambda=0$ limit.
This involves an action of $\Gsix$ on the $t,z,\rho$ and $\text{S}^3$ directions.
This $\Gsix$ transformation should be viewed as a solution generating transformation rather than a U-duality.
It may at first seem highly mysterious that the group $\Gsix$ should appear rather than the $\Gfour$ we used to generate the solution: this can be explained by tracing the origin of this deformation back to an $\mathrm{SO}(2,2)$ T-duality transformation acting just on the $(t,z)$ directions of the original F1-NS5 solution. Our full solution therefore inherits non-trivial structure associated to the action of `duality' transformations in $2+4=6$ directions, which singles out $\Gsix$. We will explain this further below.

An $\Gsix$ transformation non-trivially mixes the metric with the three-form and six-form potentials, which can be explicitly introduced as:
\be
\begin{split}
C_{(3)} & = \frac{r^2 \rho }{1+\lambda r^2} \dd t \wedge \dd z \wedge \dd \rho+ \frac{\rho^4}{1+\lambda r^2 + \rho^2} \mathrm{Vol}_{\text{S}^3} \,,\\
C_{(6)} & = \rpm \frac{ r^2 \rho^3}{2} \left(\frac{1}{1+\lambda r^2} + \frac{1}{1+\lambda r^2 +\rho^2} \right) \dd t \wedge \dd z \wedge \dd \rho \wedge \mathrm{Vol}_{\text{S}^3}\,.
\end{split} 
\label{sixpots}
\ee
The remaining components of $C_{(3)}$ and $C_{(6)}$, which have components along the torus, are electromagnetically dual to those written here.
The relevant component of the dual field strength leading to the six-form potential is
\be
\begin{split}
\star F_{(4)} & \supset
\rpm \frac{2  \rho^3 r\dd t \wedge \dd z \wedge \dd r \wedge \dd \rho \wedge \mathrm{Vol}_{\text{S}^3}}{(1+\lambda r^2)(1+\lambda r^2+ \rho^2)} 
\end{split} 
\label{dualFsolnD}
\ee
As $d \star F_{(4)} \rpm \tfrac{1}{2} F_{(4)} \wedge F_{(4)}=0$ we then define $C_{(6)}$ by $d C_{(6)} = \star F_{(4)} \rpm \tfrac{1}{2} C_{(3)} \wedge F_{(4)}$.
The gauge choice for $C_{(6)}$ has been chosen so that it is finite for $\lambda \rightarrow 0$.

To describe the action of $\Gsix$, we make a $(6+5)$-dimensional split of the coordinates.
Let $x^{\ii} = (t,z,\rho,\theta^\alpha)$, where $\theta^\alpha$ denote the coordinates on the unit sphere, and let $x^\mu = (r, y^1, \dots, y^4)$ with the $y^i$ corresponding to the torus coordinates.
We decompose the metric as
\be
\\ ds^2 = \phi_{\ii \jj} \dd x^{\ii} \dd x^{\jj} + |\phi|^{-1/3} g_{\mu\nu} \dd x^\mu \dd x^\nu \,,
\ee
such that the metric $g_{\mu\nu}$ is an $\Gsix$ invariant given by
\be
g_{\mu\nu} \dd x^\mu \dd x^\nu = r^{4/3} \rho^2 (\det g_{\text{S}^3})^{1/3} \left( \frac{\dd r^2}{r^2} + \dd s^2_{\text{T}^4} \right) \,.
\ee
In particular, it is independent of $\lambda$. 

The metric $\phi_{\ii \jj}$ transforms alongside the three-form components $C_{\ii\jj\kk}$ and the six-form component $C_{\ii\jj\kk\ll\mm\nn}\equiv  C \epsilon_{\ii\jj\kk\ll\mm\nn}$.
The $\Gsix$ covariant object containing these fields is a 27 $\times$ 27 generalised metric. 
This can be written as \cite{Berman:2011jh, Lee:2016qwn}
\be
\gM_{MN}(\phi,C_{(3)},  C_{(6)}) = U_M{}^K\bar\gM_{KL} U_N{}^L \,,\quad
\bar\gM_{MN} = |\phi|^{1/3}
\begin{pmatrix}
\phi_{\ii\jj} & 0 & 0 \\
0 & 2 \phi^{\ii[\jj} \phi^{\jj']\ii'} & 0\\
0 & 0 & (\det \phi)^{-1} \phi_{\ii\jj} 
\end{pmatrix} \,,
\label{defE6GM}
\ee
\be
U_M{}^N = \begin{pmatrix}
\delta_{\ii}{}^{\jj} &- C_{\ii \jj\jj'} & \rmp \delta_{\ii}{}^{\jj}  C + \tfrac{1}{4!} \epsilon^{\jj \kk_1 \dots \kk_5} C_{\ii\kk_1\kk_2} C_{\kk_3 \kk_4\kk_5} \\
0 & 2 \delta^{\ii\ii'}_{\jj\jj'} & - \tfrac{1}{3!} \epsilon^{\ii\ii' \jj \kk_1 \kk_2 \kk_3} C_{\kk_1 \kk_2 \kk_3} \\
0 & 0 & \delta_{\ii}{}^{\jj}
\end{pmatrix}\,.
\ee
Here the 27-dimensional $\Gsix$ fundamental index decomposes as $V^M = ( V^{\ii}, V_{\ii\ii'}, V^{\bar{\ii}})$ where $V_{\ii\ii'} = - V_{\ii'\ii}$ and $V^M W_M \equiv V^{\ii} W_{\ii} + \tfrac{1}{2} V_{\ii\ii'} W^{\ii\ii'} + V^{\bar{\ii}} W_{\bar{\ii}}$. There are thus two six-dimensional vector indices: the second one can be viewed as coming from a dualisation of five-form indices $V^{\bar {\ii}} \equiv \tfrac{1}{5!} \epsilon^{{\ii} {\jj}_1 \dots \jj_5} V_{\jj_1 \dots \jj_5}$.\footnote{Here both $\epsilon^{012345} = \epsilon_{012345} = +1$ are Levi-Civita symbols defined without relative minus signs for convenience.}

It is straightforward to evaluate the generalised metric for the six-dimensional metric and form-fields obtained from \eqref{solutionDeformation}.
Some general formulae applicable to situations where the six-dimensional metric and form-fields admit a (3+3)-dimensional decomposition are recorded in appendix \ref{e6technology}.
One finds that the generalised metric depends \emph{linearly} on $\lambda$, and furthermore that the $\lambda$ dependence can be factorised via an $\Gsix$-valued transformation involving a six-vector parameter.
Generally, we can introduce an $\Gsix$-valued matrix describing deformations involving a trivector $\Omega^{\ii\jj\kk}$ and a six-vector $\Omega^{\ii\jj\kk\ll\mm\nn} \equiv \Omega \epsilon^{\ii\jj\kk\ll\mm\nn}$, such that \cite{Lee:2016qwn}
\be
\widetilde U_M{}^N = \begin{pmatrix}
 \delta_{\ii}{}^{\jj} &0 & 0  \\
- \Omega^{\ii\ii' \jj} & 2 \delta^{\ii\ii'}_{\jj\jj'} & 0\\
\delta_{\ii}{}^{\jj} \Omega +  \tfrac{1}{4!} \epsilon_{\ii \kk_1 \dots \kk_5} \Omega^{j\kk_1\kk_2} \Omega^{\kk_3 \kk_4\kk_5}  & - \tfrac{1}{3!} \epsilon_{\ii \jj\jj' \kk_1 \kk_2 \kk_3} \Omega^{\kk_1 \kk_2 \kk_3}  & \delta_\ii{}^\jj
\end{pmatrix}\,.
\label{defUOmega}
\ee
Again using the formulae in appendix \ref{e6technology}, it can be straightforwardly checked that the generalised metric describing the background \eqref{solutionDeformation} admits a factorisation
\be
\gM_{M N} (\lambda) = \widetilde U_M{}^K(\lambda) \gM_{KL} (\lambda=0) \widetilde U_N{}^L(\lambda)
\label{nicefactorisation}
\ee
where $\widetilde U_M{}^N(\lambda)$ has the form of \eqref{defUOmega} with 
\be
\Omega^{\ii\jj\kk}=0\,,\quad
 \Omega =  \rpm \frac{\lambda}{2 \rho^3 \sqrt{\det g_{\text{S}^3}}}
\,,
\ee
where $\sqrt{\det g_{\text{S}^3}}$ denotes the volume element on the unit three-sphere. 
Hence the factorisation \eqref{nicefactorisation} demonstrates that the full solution \eqref{solutionDeformation} is a six-vector deformation of the $\lambda=0$ background corresponding to the AdS limit.

The fact that the deformation parameter is non-constant can be understood by viewing this form of the deformation as involving a change of coordinates as well as a constant $\Gsix$ transformation.
This change of coordinates is just that which defines Cartesian coordinates $x^i$ in place of the `spherical' coordinates $(\rho, \theta^\alpha)$.
In terms of the Cartesian coordinates one has simply: 
\be
\Omega^{tz ijkl} = \rpm \frac{\lambda}{2} \epsilon^{ijkl}\,.
\ee
It is still non-trivial that this is a solution generating transformation, as the full solution depends on the $x^i$ coordinates, and so we are not in a situation with isometries to which we would automatically be entitled to apply U-duality transformations.
The six-vector deformation however commutes with the EDA generalised frame containing the trivector $\Omega^{ijk} \sim \epsilon^{ijkl} x_l$.
Prior to applying the EDA generalised frame, what we have is an 11-dimensional configuration (that is not a solution) which already admits the six-vector factorisation.

This follows directly from the properties of the original F1-NS5 extremal solution.
Using the same coordinate redefinition that introduces the parameter $\lambda$, the F1-NS5 extremal solution \eqref{extremalF1NS5} can be written as\footnote{This rewriting is inspired by \cite{Giveon:2017nie,Aguilera-Damia:2020qzw}.}
\be
\begin{split}
ds_s^2 & = \frac{r^2}{1+\lambda r^2} ( -  \dd t^2 + \dd z^2 )+  \frac{\dd r^2}{r^2} +  	  \dd s^2_{\text{S}^3} + \dd s^2_{\text{T}^4} \,,\quad
B_{tz} = \frac{r^2 }{1+\lambda r^2}  \,,\quad e^{-2\varphi}  = 1+\lambda r^2  \,.
\end{split} 
\label{extremalF1NS5again}
\ee
The $\lambda$ dependence now corresponds to an $\mathrm{SO}(2,2)$ T-duality deformation acting on the $(t,z)$ directions.
This is seen by passing to the appropriate $\mathrm{SO}(2,2)$ covariant description via a generalised metric
\be
\cH_{MN}(\lambda)  = \begin{pmatrix} 
g-Bg^{-1}B &  B g^{-1} \\
- g^{-1} B & g^{-1}
\end{pmatrix}
= \begin{pmatrix} 0 & Z \\
Z & ( r^{-2} + \lambda ) \eta
\end{pmatrix} 
\,,\quad Z \equiv \begin{pmatrix} 0 & 1 \\ 1 & 0 \end{pmatrix} \,,\quad
\eta \equiv \begin{pmatrix} -1 & 0 \\ 0 & 1 \end{pmatrix} \,,
\ee
factorising as 
\be
\cH_{MN}(\lambda) = U_M{}^K(\lambda) \cH_{KL}(\lambda=0) U_N{}^L(\lambda) \,,\quad
U_M{}^N = \begin{pmatrix} 1 & 0 \\
-\beta & 1 \end{pmatrix} \,,\quad
\beta \equiv \frac{\lambda}{2} \begin{pmatrix} 0 & 1 \\ -1 & 0 \end{pmatrix} \,.
\ee
The deformation matrix $\beta$ has an interpretation as a bivector $\beta^{ij}$. 
(This can alternatively be seen as a TsT transformation.)
In addition, the $\mathrm{SO}(2,2)$ invariant generalised dilaton is $e^{-2\varphi}\sqrt{|\det(g)|}=r^2$ and is independent of $\lambda$.

When we apply the reduction ansatz for type IIA on $\text{S}^3$ to the F1-NS5 background, the field strength component $H_{tzr}=\partial_r B_{tz}$ becomes the $\fA = 5$ component of the $\Gfour$ covariant field strength $\cH_{(3) \fA}$.
On uplifting to an eleven-dimensional solution (using the coordinates $x^i$), this leads to the identification $F_{tzr ijkl} \sim H_{tzr} \epsilon_{ijkl}$ giving a non-trivial dual seven-form field strength.
Hence the B-field component $B_{tz}$ induces the component $C_{tzijkl}$ of the eleven-dimensional dual six-form.
Accordingly, the bivector deformation $\beta^{tz}$ becomes the six-vector deformation $\Omega^{tz ijkl} =\beta^{tz} \epsilon^{ijkl}$.
The smallest U-duality group capable of describing such a deformation is $\Gsix$, and this provides the exact explanation for why $\Gsix$ appears.

The structure of the F1-NS5 solution appearing here is associated to some intriguing physics.
The solution can be viewed as interpolating from an AdS${}_3$ geometry to a linear dilaton spacetime, holographically dual to Little String Theory \cite{Aharony:1998ub,Giveon:1999zm}.
This interpolation, realised above via the bivector deformation, has been argued to correspond to a single-trace $T \bar T$ deformation of the dual CFT${}_2$ \cite{Giveon:2017nie}, and has a worldsheet interpretation as a marginal current-current coupling.
We might therefore expect that our full solution captures again a deformation related to $T \bar T$ of the CFTs dual to the AdS${}_3$ limit of our solution (these are the quiver field theories described in \cite{Lozano:2020bxo}).
Making this precise would be interesting future work.

A final comment here is that deformations of the form \eqref{defUOmega} generically lead to terms quadratic in the six-vector deformation unless the upper left block of the generalised metric vanishes, $\gM_{\ii\jj}=0$.
This block is of the form $\gM_{\ii \jj} \sim (\phi + C_{(3)}^2 + ( C_{(6)} + C_{(3)}^2)^2 )_{\ii \jj}$ and so involves terms quadratic $C_{(6)}$ as well as both quadratic and quartic in $C_{(3)}$.
Rather remarkably the gauge choice made above for the three- and six-form is such that here $\gM_{\ii\jj} = 0$.

\subsection{Supersymmetry}
\label{susy}

In this section we discuss the supersymmetry of the AdS${}_3$ limit \eqref{AdSlimitSimpler} of our solution.
The Killing spinor equation in our conventions\footnote{We follow \cite{Ortin:2004ms} so that $\{ \Gamma_a , \Gamma_b \} = 2 \eta_{ab}$ with $\eta_{ab}$ having mostly minus signature.}
\begin{equation}
\delta_{\epsilon} \psi_\mu=
2D_{\mu}\epsilon
+\tfrac{i}{144}(\Gamma^{\nu\rho\sigma\lambda}{}_{\mu}-8\Gamma^{\rho\sigma\lambda}\delta^{\nu}_{\mu})\epsilon F_{\nu\rho\sigma\lambda}=0\,.
\label{KS}
\end{equation}
We will proceed to solve this explicitly, finding a $\tfrac12$-BPS solution \eqref{fullsolution}.
We denote the AdS coordinates by $(t,z,r)$, the torus coordinates by $y^i$, $i=1,\dots,4$ and the (standard) three-sphere coordinates by $(\chi, \theta,\varphi)$.
Unless otherwise indicated, in the below equations the indices on the gamma matrices should be assumed to be flat.

We first assume that $\epsilon$ is independent of the torus coordinates $y^i$.
Then the $\mu=y^i$ components of \eqref{KS} provide an algebraic condition on $\epsilon$: 
\begin{equation}\label{yy}
\Big[\rho(1+\rho^{2})^{-1}\Gamma_{\rho }-\frac{i}{2}(1+\rho^{2})^{-1/2}\Big(2\rho\Gamma^{tzr\rho}-4\Gamma^{y_{1}...y_{4}}+4(1+\frac{1}{2}\rho^{2})(1+\rho^{2})^{-1/2}\Gamma^{\rho\chi\theta\varphi}\Big)\Big]\epsilon=0
\,.
\end{equation}
The AdS components of \eqref{KS} give differential equations
\be
D_{\hat m} \epsilon + \tfrac{1}{6} \Gamma_{\hat m} X \epsilon = 0 \,,
\label{KSADS}
\ee
where
\be
X = \left(
- (1+\rho^2)^{-1} \rho \Gamma_{\rho} + i (1+\rho^2)^{-1/2} \left(
- 2 \rho \Gamma^{t z r \rho} - \Gamma^{y_1 \dots y_4} + 2 ( 1 + \tfrac12 \rho^2 ) (1+\rho^2)^{-1/2} \Gamma^{\rho\chi\theta\varphi} 
\right)
\right) \,.
\ee
In \eqref{KSADS} $\hat m$ denote curved AdS indices.
The spin connection components are $D_{\hat r} \epsilon = \partial_r \epsilon$ and $D_{\hat a} \epsilon = \partial_a \epsilon -\tfrac12 \Gamma_{a r} \epsilon$, with $\hat a$ labelling the $t$ and $z$ directions, and $\Gamma_{\hat r} = r^{-1} \Gamma_{r}$, $\Gamma_{\hat a} = r \Gamma_a$ where $\Gamma_r$ and $\Gamma_a$ are the gamma matrices with flat indices.
The form of the $r$-dependence of the $\hat m = r$ equation implies that the $r$-dependence of $\epsilon$ has to be of the form $r^{\beta}$, with a matrix $\beta$ to be determined later, leading to a further algebraic condition on $\epsilon$.
Indeed, letting explicitly $\epsilon = r^\beta \tilde \epsilon$, where $\tilde \epsilon$ depends on $t,z$ and the other spacetime coordinates, we get an equation
\be
( \beta + \tfrac16 \Gamma_r X ) \epsilon = 0 \,.
\label{adsalg}
\ee
It follows that $D_{\hat m} \epsilon = - \Gamma_{\hat m} \Gamma_r \beta \epsilon$.
For the $(t,z)$ components we get
\be
\partial_a \epsilon = \Gamma_{ar} ( -\beta + \tfrac{1}{2} ) \epsilon 
\Rightarrow 
\partial_a \tilde \epsilon = r^{-\beta} \Gamma_{ar} r^{\beta} \tfrac{1}{2} ( 1 - 2 \beta) \tilde \epsilon \,.
\label{tildeeps}
\ee
We have an $r$-independent expression on the left hand side, and so by our assumptions the right hand side of has to be $r$-independent as well, thus, differentiating the right hand side with respect to $r$ we end up requiring the following expression to vanish:
\begin{equation}
r^{-\beta}\Big(\Gamma_{ar}-[\beta,\Gamma_{ar}]\Big)(1-2\beta)r^{\beta}\tilde{\epsilon}=0\,,
\end{equation}
which can be achieved if 
\begin{equation}
\Big(\Gamma_{ar}-[\beta,\Gamma_{ar}]\Big)(1-2\beta)=0\,.
\end{equation}
If $\beta$ commutes with $\Gamma_{ar}$ then the only solution is $\beta = \tfrac{1}{2} I$.
Alternatively, if $\beta$ anticommutes with $\Gamma_{ar}$, then we can extract $\Gamma_{ar}$ from the equation again leading to 
\begin{equation}
\Big((2\beta)^{2}-1\Big)=0\,,
\end{equation}
which tells us $2\beta$ should square to a unit matrix.
This condition and that of anticommuting with $\Gamma_{tr}$ and $\Gamma_{zr}$ is compatible with multiple choices for $\beta$, for instance $2\beta=\pm\Gamma_{tz}$, $2\beta=\pm i\Gamma_{r}$, $2\beta=\pm \Gamma_{tzy_{1}...y_{4}}$. However, not all options will lead to a non-trivial solution for $\tilde \epsilon$, and some of them have fewer supersymmetries than others, as we will see shortly.

Now let's assemble and make sense of the algebraic conditions on $\epsilon$. 
We can rewrite \eqref{adsalg} as
\begin{equation}\label{rr}
\Big[2\Gamma_{r}\beta+\frac{1}{3}\rho(1+\rho^{2})^{-1}\Gamma_{\rho}-\frac{i}{6}(1+\rho^{2})^{-1/2}\Big(4(1+\frac{1}{2}\rho^{2})(1+\rho^{2})^{-1/2}\Gamma^{\rho\chi\theta\varphi}-2\Gamma^{y_{1}...y_{4}}-4\rho\Gamma^{tzr\rho}\Big)\Big]\epsilon=0
\end{equation}
Subtracting $\frac{1}{3}$ (\ref{yy}) from (\ref{rr}) we get:
\begin{equation}\label{cc1}
\Big[2\Gamma_{r}\beta+i(1+\rho^{2})^{-1/2}\big(\rho\Gamma^{tzr\rho}+\Gamma^{y_{1}...y_{4}}\big)\Big]\epsilon=0
\end{equation}
This (for suitable $\beta$) will provide a coordinate-dependent projector condition on $\epsilon$, similar to that appearing in non-abelian T-dual solutions \cite{Sfetsos:2010uq}.
We can also deduce a second projector condition. Let's first split the $\Gamma^{\rho\chi\theta\varphi}$ and $\Gamma_{r}$ parts of \eqref{rr} as
\be
\begin{split}
\frac{1}{3}& \Big[2\Gamma_{r}\beta+i(1+\rho^{2})^{-1/2}\Big(\rho\Gamma^{tzr\rho}+\Gamma^{y_{1}...y_{4}}\Big)\Big]\epsilon 
\\ & +\frac{2}{3}\Big[2\Gamma_{r}\beta+\rho(1+\rho^{2})^{-1}\Gamma_{\rho} 
-i\Gamma^{\rho\chi\theta\varphi}-i(1+\rho^{2})^{-1}\Gamma^{\rho\chi\theta\varphi}-i(1+\rho^{2})^{-1/2}\Gamma^{y_{1}...y_{4}}\Big]\epsilon=0
\end{split}
\ee
the first line of which is exactly (\ref{cc1}) thus vanishes. We can write the second line as 
\begin{equation}
\Big[2\Gamma_{r}\beta-i\Gamma^{\rho\chi\theta\varphi}+\Gamma^{tzr}(1+\rho^{2})^{-1}\Big(\rho\Gamma^{tzr\rho}+i\Gamma^{tzr\rho\chi\theta\varphi}\Big)-i(1+\rho^{2})^{-1/2}\Gamma^{y_{1}...y_{4}}\Big]\epsilon=0
\end{equation}
then using the fact that the product of all gamma matrices is (in our conventions) $-i$,
 we can rewrite $\Gamma^{tzr\rho\chi\theta\varphi}=i\Gamma^{y_{1}...y_{4}}$, and use (\ref{cc1}) again to obtain
\begin{equation}
\Big[2\Gamma_{r}\beta-i\Gamma^{\rho\chi\theta\varphi}+i(1+\rho^{2})^{-1/2}\Big(\Gamma^{tz}-\Gamma^{y_{1}...y_{4}}\Big)\Big]\epsilon=0
\end{equation}
and then again rewriting $\Gamma^{y_{1}...y_{4}}=i\Gamma^{tzr}\Gamma^{\rho\chi\theta\varphi}$, and $\Gamma^{tz}=-\Gamma^{tzr}\Gamma^{r}$, we finally extract a common factor
\begin{equation}
\Big(1+i(1+\rho^{2})^{-1/2}\Gamma^{tzr}\Big)\Big[2\Gamma_{r}\beta-i\Gamma^{\rho\chi\theta\varphi}\Big]\epsilon=0
\end{equation}
multiplying this by $\Big(1-i(1+\rho^{2})^{-1/2}\Gamma^{tzr}\Big)$ and extracting the non-negative resulting $\rho^{2}$ we arrive at the second projector condition on $\epsilon$:
\begin{equation}\label{cc2}
\Big[2\Gamma_{r}\beta-i\Gamma^{\rho\chi\theta\varphi}\Big]\epsilon=0\,.
\end{equation}
As we want our solution to be as supersymmetric as possible, we want to choose a $\beta$ that will cancel some of the algebraic conditions on $\epsilon$. Looking at (\ref{cc2}) and keeping in mind that $\Gamma^{tzry_{1}...y_{4}}=-i\Gamma^{\rho\chi\theta\varphi}$, we immediately see that the choice $2\beta=\Gamma^{tzy_{1}...y_{4}}$ will turn this condition into a trivial one! Thus, we can conclude the choice $2\beta=\Gamma^{tzy_{1}...y_{4}}$ corresponds to a most supersymmetric solution; other choices would impose \eqref{cc2} and lead to a solution with fewer supersymmetries.

Now let us look at the full AdS part of the solution that corresponds to $2\beta=\Gamma^{tzy_{1}...y_{4}}$ and then come back to the remaining equations.
We will write our solution in the form
\begin{equation}
\epsilon=\epsilon_{\text{AdS}}\epsilon_{\rho}\epsilon_{\text{S}^3}\epsilon_{0}
\end{equation}
with $\epsilon_{0}$ is a constant spinor and the other factors are matrices depending on the AdS, $\rho$ and sphere coordinates respectively.

The differential equation \eqref{tildeeps} on $\tilde{\epsilon}$ becomes
\begin{equation}
\partial_{a}\tilde{\epsilon}=\frac{1}{2}\Gamma_{ar}(1-2\beta)\tilde{\epsilon}
\end{equation}
with the solution
\begin{equation}
\tilde{\epsilon}=\exp\Big[\frac{1}{2}x^{a}\Gamma_{ar}(1-2\beta)\Big] \bar \epsilon
=\left(1+\frac{1}{2}x^{a}\Gamma_{ar}(1-2\beta) \right) \bar \epsilon
\end{equation}
where in the second equality we take into account our previous assumption that $\beta$ anticommutes with $\Gamma_{ar}$ and $(2\beta)^{2}=I$ so that we can make an expansion of the exponent to the linear term.
Here $\bar \epsilon = \epsilon_{\rho}\epsilon_{\text{S}^3}\epsilon_{0}$.
Hence the full factor $\epsilon_{\text{AdS}}$ is
\begin{equation}
\epsilon_{\text{AdS}}=r^{\frac{1}{2}\Gamma^{tzy_{1}...y_{4}}}\Big(1+\frac{1}{2}x^{a}\Gamma_{ar}(1-\Gamma^{tzy_{1}...y_{4}})\Big)\,.
\end{equation}
Expanding the $r$ exponent, this can be seen to match the form of the AdS solutions obtained in \cite{Lu:1998nu}.

Now we consider the remaining differential equations on $\epsilon$. 
We start with the case corresponding to the $\rho$ coordinate:
\begin{equation}
\partial_{{\rho}}\epsilon-\frac{i}{6}(1+\rho^{2})^{-1/2}\Gamma_{\rho}\Big[\Gamma^{y_{1}...y_{4}}+2\rho\Gamma^{tzr\rho}+4(1+\frac{1}{2}\rho^{2})(1+\rho^{2})^{-1/2}\Gamma^{\rho\chi\theta\varphi}\Big]\epsilon=0
\label{KSrho}
\end{equation}
Using the projector conditions \eqref{cc1} and \eqref{cc2} (the latter of course now an identity given the form of $\beta$), as well as gamma matrix identities, we can simplify this to
\begin{equation}
\partial_{{\rho}}\epsilon-\frac{1}{6}\rho(1+\rho^{2})^{-1}\epsilon+\Gamma_{r\rho}\beta(1+\rho^{2})^{-1}\epsilon=0\,.
\end{equation}
and now the solution for $\epsilon_{\rho}$ will depend on how $\epsilon_{\text{AdS}}$ permutes with $\beta$.
For our choice of $\beta$, all the matrices in $\epsilon_{\text{AdS}}$ commute with $\Gamma_{r\rho}\beta$, and we can simply move $\epsilon_{\text{AdS}}$ to the left of each term in the equation.
We then end up with a differential equation for $\epsilon_\rho$ with the following solution:
\begin{equation}
\epsilon_{\rho}=(1+\rho^{2})^{1/12}\exp\Big[\frac{1}{2}\tan^{-1}\rho \Gamma_{tzry_{1}...y_{4}\rho}\Big]
\label{epsilonrho}
\end{equation}
We move on to the sphere components of the Killing spinor equation. We let $\epsilon_{\text{S}^3} = \epsilon_\chi(\chi) \epsilon_\theta(\theta) \epsilon_{\varphi} (\varphi)$.
The $\chi$ equation becomes after similar simplifications using the projector conditions
\begin{equation}
\partial_{\chi}\epsilon+\frac{1}{2}(1+\rho^{2})^{-1/2}\Gamma_{\rho\chi}\Big[1+\rho\Gamma_{\rho tzry_{1}...y_{4}}\Big]\epsilon=0\,,
\label{KSchi}
\end{equation}
or
\begin{equation}
\partial_{\chi}\epsilon+\frac{1}{2}\exp[\Gamma_{tzry_{1}...y_{4}\rho}\tan^{-1}\rho]\Gamma_{\rho\chi}\epsilon=0\,.
\end{equation}
Permuting $\Gamma_{\rho\chi}$ in the second term in this equation with $\epsilon_{\rho}$ we change the sign in the exponent of $\epsilon_{\rho}$ from equation \eqref{epsilonrho}, which combined with the exponential of this equation gives the same $\epsilon_{\rho}$ finally in the second term on the left. Thus, after extracting $\epsilon_{\rho}$ from the both terms of the equation to the left, we have the simple equation
\begin{equation}
\partial_{\chi}\epsilon_{\chi}+\frac{1}{2}\Gamma_{\rho\chi}\epsilon_{\chi}=0
\Rightarrow
\epsilon_{\chi}=\exp\Big[-\frac{1}{2}\Gamma_{\rho\chi}\chi\Big]\,.
\end{equation}
The same technique can be applied to obtain $\epsilon_{\theta}$ and $\epsilon_{\varphi}$ parts of the solution, which end up being
\begin{equation}
\epsilon_{\theta}=\exp\Big[-\frac{1}{2}\Gamma_{\chi\theta}\theta\Big] \,,\quad
\epsilon_{\varphi}=\exp\Big[-\frac{1}{2}\Gamma_{\theta\varphi}\varphi\Big]\,.
\end{equation}
The full solution we have obtained can therefore be written as
\begin{equation}
\epsilon=(1+\rho^{2})^{1/12}r^{\beta}\Big(1+\frac{1}{2}x^{a}\Gamma_{ar}(1-2\beta)\Big)\exp\Big[-\beta\Gamma_{r\rho}\tan^{-1}\rho\Big]\epsilon_{\Omega}\epsilon_{0}\,,
\label{fullsolution}
\end{equation}
with $\beta=\frac{1}{2}\Gamma^{tzy_{1}...y_{4}}$, $\epsilon_{\Omega}=\epsilon_{\chi}\epsilon_{\theta}\epsilon_{\varphi}$.
In addition we have the projector condition \eqref{cc1}, which we can rewrite as
\be
\left( 1 + \tfrac{i}{\sqrt{1+\rho^2}} \left(  \Gamma^{tzr} - \rho \Gamma^{y_1 \dots y_4 \rho}  \right) \right) \epsilon = 0 \,.
\ee
This can be shown to reduce to a single projector condition on the constant spinor $\epsilon_0$.
To show this, we apply the projector condition in its original form \eqref{cc1} to \eqref{fullsolution} and proceed as follows.
We first permute the exponential in $\epsilon_\rho$ with $\Gamma_{r}\beta$ from (\ref{cc1}). After then factoring out a common $\epsilon_{\rho}$ we can use the identities $\sin\tan^{-1}\rho=\rho(1+\rho^{2})^{-1/2}$, $\cos\tan^{-1}\rho=(1+\rho^{2})^{-1/2}$ to rewrite \eqref{cc1} applied to \eqref{fullsolution} as
\begin{equation}
(1+\rho^{2})^{-1}\Big[\big(1+\rho\Gamma_{tzry_{1}...y_{4}\rho}\big) 2\Gamma_{r}\beta+i\big(\rho\Gamma^{tzr\rho}+\Gamma^{y_{1}...y_{4}}\big)\Big]\epsilon_{\chi}\epsilon_{\theta}\epsilon_{\varphi}\epsilon_{0}=0\,.
\end{equation}
Then permuting with $\epsilon_{\chi}$, the terms linear in $\rho$ give different signs in the exponent containing $\Gamma_{\rho}$, leading to 2 equations:
\begin{equation}
(2\Gamma_{r}\beta+i\Gamma^{y_{1}...y_{4}})\epsilon_{0}=0
\,,\quad (\Gamma_{\rho}+i\Gamma^{tzr\rho})\epsilon_{0}=0\,.
\end{equation}
However these are actually equivalent and give the single condition:
\begin{equation}
(1+i\Gamma^{tzr})\epsilon_{0}=0\,.
\end{equation}
Therefore we have 1 condition on $\epsilon_{0}$, reducing the degrees of freedom by $\frac{1}{2}$, so this is a $\frac{1}{2}$-BPS solution.
This is the same amount of supersymetry as the original F1-NS5 solution in its AdS${}_3$ limit.
Away from this limit we expect our full solution \eqref{solnSphericalSimpler} is $\tfrac14$-BPS.
It is worth noting that the solutions of \cite{Lozano:2020bxo} are generically $\tfrac14$-BPS, suggesting that our solution allows for an enhancement, likely due to the special case $k=1$.
We note that a similar explicit Killing spinor solution was found in \cite{Zacarias:2021pfz}.

\subsection{IIA reductions}
\label{reductions}

Finally, let us record the expressions for different solutions of type IIA supergravity which can be obtained by reducing the solution \eqref{solnSphericalSimpler} in different ways.
All these solutions could further be T-dualised in multiple ways to give solutions of type IIB supergravity.

\paragraph{Reduction on $\text{T}^4$ direction}

Reducing on one of the $\text{T}^4$ directions we obtain
\be
\begin{split} 
\dd s_{10}^2  &= (r^2 f_1 + \rho^2)^{1/2}  (r^2 f_1)^{1/2} 
\left(  
\frac{1}{f_1} ( - \dd t^2 + \dd z^2 )
+ \frac{\dd r^2}{r^2} 
+ \frac{\dd \rho^2}{r^2 f_1} 
+ \dd s^2_{\text{T}^3}
\right)
\\ & \quad
 + ( r^2 f_1 +  \rho^2)^{-1/2} (r^2f_1)^{1/2} \rho^2 \dd s^2_{\text{S}^3} 
\,,
\\
H_{(3)} & = \rpm 2 \mathrm{Vol}_{\text{T}^3} \,,\quad e^{-2\varphi} = (r^2 f_1+\rho^2)^{-1/2} (r^2f_1)^{-1/2} \,, \quad F_{(2)} = 0 \,,
\\
F_{(4)} & = 
\frac{2 r \rho }{(r^2 f_1)^2}\dd t \wedge \dd z \wedge \dd r \wedge \dd\rho 
+ \frac{( 4 r^2 f_1 + 2  \rho^2 ) }{(r^2 f_1 +  \rho^2 )^2}  \rho^3 \dd\rho \wedge \mathrm{Vol}_{\text{S}^3} 
- \frac{\rho^4 \partial_r ( r^2 f_1)}{(r^2 f_1 +  \rho^2)^2}   \dd r \wedge \mathrm{Vol}_{\text{S}^3}  \,.
\end{split} 
\label{solnIIA1}
\ee 
This still has an AdS${}_3$ near horizon limit, and the full solution is a six-vector deformation of this. The six-vector is now associated to the NSNS six-form.

\paragraph{Reduction on Hopf fibre}
Writing the metric on the three-sphere as
\be
\dd s^2_{\text{S}^3}=\frac{1}{4}\left( ( \dd \psi+\eta)^2+\dd s^2_{\text{S}^2}\right)\,,\quad \dd \eta={\text{Vol}}_{\text{S}^2}\,.
\label{hopf}
\ee
and reducing on the Hopf fibre direction parametrised by $\psi$ we obtain
\be
\begin{split} 
\dd s_{10}^2  &= (r^2 f_1)^{1/2} \frac{\rho}{2} 
\left(  
\frac{1}{f_1} ( - \dd t^2 + \dd z^2 ) +    \frac{\dd r^2}{r^2}
+ \frac{\dd \rho^2}{r^2 f_1} 
 + \dd s^2_{\text{T}^4}
\right)
\\ & \quad
 + ( r^2 f_1 +  \rho^2)^{-1} (r^2f_1)^{1/2} \left(\frac{\rho}{2}\right)^3 \dd s^2_{\text{S}^2} 
\,,
\\
H_{(3)} & = \tfrac18 \frac{( 4 r^2 f_1 + 2  \rho^2 ) }{(r^2 f_1 +  \rho^2 )^2}  \rho^3 \dd\rho \wedge \mathrm{Vol}_{\text{S}^2} 
- \tfrac18 \frac{\rho^4 \partial_r ( r^2 f_1)}{(r^2 f_1 +  \rho^2)^2}   \dd r \wedge \mathrm{Vol}_{\text{S}^2} \,,\\
 e^{-2\varphi} & = (r^2f_1+\rho^2) (r^2f_1)^{-1/2} \left(\tfrac{\rho}{2}\right)^{-3} \,,\\
F_{(2)} & = \mathrm{Vol}_{\text{S}^2} \,,\quad F_{(4)}  = 
\frac{2 r \rho }{(r^2 f_1)^2}\dd t \wedge \dd z \wedge \dd r \wedge \dd\rho 
\rpm 2 \mathrm{Vol}_{\text{T}^4}\,.
\end{split} 
\label{solnIIA2}
\ee 
This still has an AdS${}_3$ near horizon limit, and the full solution is a five-vector deformation of this, with the five-vector associated to the RR five-form.
As the M-theory AdS${}_3 \times \text{S}^3$ solutions of \cite{Lozano:2020bxo} were obtained by uplifting the AdS${}_3 \times \text{S}^2$ IIA solutions constructed in \cite{Lozano:2019emq,Lozano:2019jza,Lozano:2019zvg,Lozano:2019ywa} on a Hopf fibre, the solution \eqref{solnIIA2} can be interpreted using the latter.

\paragraph{Reduction on AdS direction} 
Reducing on the $z$ direction we obtain
\be
\begin{split} 
\dd s_{11}^2  &= (r^2 f_1 + \rho^2)^{1/2}  r 
\left(  
- \frac{1}{f_1}\dd t^2
+    \frac{\dd r^2}{r^2}
+ \frac{\dd \rho^2}{r^2 f_1} 
 + \dd s^2_{\text{T}^4}
\right)
 + ( r^2 f_1 +  \rho^2)^{-1/2} r \rho^2 \dd s^2_{\text{S}^3} 
\,,
\\
H_{(3)} & = \frac{2 r \rho }{(r^2 f_1)^2}\dd t \wedge \wedge \dd r \wedge \dd\rho \,,\quad e^{-2\varphi} = (r^2 f_1+\rho^2)^{-1/2} \tfrac{f_1}{r} \,,\quad F_{(2)}  = 0 \,,
\\
F_{(4)} & = 
\rpm 2 \mathrm{Vol}_{\text{T}^4}
+\frac{( 4 r^2 f_1 + 2  \rho^2 ) }{(r^2 f_1 +  \rho^2 )^2}  \rho^3 \dd\rho \wedge \mathrm{Vol}_{\text{S}^3} 
- \frac{\rho^4 \partial_r ( r^2 f_1)}{(r^2 f_1 +  \rho^2)^2}   \dd r \wedge \mathrm{Vol}_{\text{S}^3}  \,.
\end{split} 
\label{solnIIA3}
\ee 
This now has an AdS$_2$ near horizon limit, and the full solution is a five-vector deformation of this. The five-vector is associated to the RR five-form.

\section{Discussion}
\label{discussion} 

In this paper we first discussed the idea of generalised T- and U-dualities, viewed as a solution generating technique in supergravity.
We reviewed how these generalised dualities can be linked to special classes of algebras, which are efficiently geometrically encoded using generalised parallelisations in generalised geometry. 
Building on our previous paper \cite{Blair:2020ndg}, we focused on an example in 11-dimensional supergravity characterised by non-vanishing dual 3-algebra structure constants in the underlying exceptional Drinfeld algebra introduced to control Poisson-Lie U-duality in \cite{Sakatani:2019zrs,Malek:2019xrf}.

To produce a new supergravity solution, we had to step slightly outside the confines of the EDA set-up.
We used the fact that our EDA generalised frame incorporating the Euclidean 3-algebra solution provided a consistent truncation to $\mathrm{CSO}(4,0,1)$ maximal gauged supergravity in 7 dimensions.
We were able to use this pragmatically to produce dual pairs of solutions by starting with the known truncation of type IIA on $\text{S}^3$ leading to the same gauged supergravity, reducing solutions of the latter form, and then uplifting with our EDA frame.
Algebraically, this alternative starting point can be viewed as relaxing the requirement that one has to pick an isotropic set of dual generators forming a subalgebra.
It would be interesting to complete this observation by formulating a more precise understanding of which families of generalised frames produce the EDA with the subalgebra requirement relaxed (the systematic approach of \cite{Inverso:2017lrz} would likely be useful here).
This would allow our construction to be viewed in terms of a slightly enlarged notion of Poisson-Lie U-duality than that initially suggested in  \cite{Sakatani:2019zrs,Malek:2019xrf}. 

The example described in this paper can be viewed as a proof-of-concept for the idea that it is possible to generate new supergravity solutions by formulating generalised notions of U-duality.
It would be beneficial to develop a more systematic approach.
For instance, it is very clear which spacetimes admit non-abelian T-duals: those with non-abelian isometries.
It is not clear what spacetimes admit generalised U-duals characterised by non-vanishing dual 3-algebra structure constants.
It is also not clear what role, if any, is played by an actual 3-algebra symmetry in such spacetimes.

Generalising to higher dimensions will also lead to higher-rank polyvectors and $n$-algebra symmetries.
It would appear that solutions characterised by an ansatz involving polyvectors linear in the coordinates have notable properties.
They describe not only the plethora of known NATD solutions, but also solutions such as the one constructed in this paper, which as we saw shared many features with solutions generated by NATD, including the general properties of its holographic completion.
Classifying and understanding the types of solutions of this form, and the possible dual solutions they may arise from, would not only help establish generalised U-duality as a useful technique on a par with non-abelian T-duality but help elucidate the general structure.

Here it would also be important to develop an understanding of which properties (supersymmetry, brane charges) of such solutions are induced by the initial solution.
For non-abelian T-duality, for example, one can precisely discuss which supersymmetries are preserved in terms of whether the action of the initial non-abelian isometries preserve the Killing spinor \cite{Sfetsos:2010uq, Itsios:2012dc, Kelekci:2014ima}.
Generically one finds a reduced amount of supersymmetry in the dual solution as a result.
In our example, in the AdS limit, we found our new solution had as many supersymmetries as the original near horizon F1-NS5 solution.
It would be useful to understand from a general viewpoint why this was the case.
This might be best formulated using exceptional field theory as a master formalism. 

It would be possible to generate further examples by focusing on specific solutions of the gauged supergravities that appear in these polyvector constructions.
For the $\mathrm{CSO}(4,0,1)$ supergravity, numerous solutions were found in \cite{Karndumri:2019osu,Karndumri:2019xqv,Karndumri:2019ejk}, all of which can be used to generate dual solutions by uplifting to type IIA on $\text{S}^3$ and to 11-dimensional supergravity via our EDA generalised frame.

Turning now to the specific example studied in this paper, this exhibits numerous interesting features linked to deformations and holographic duality. 
We argued that a holographic completion of the AdS${}_3$ limit of our solution can be obtained from the class of solutions obtained in \cite{Lozano:2020bxo}, which have well-defined quiver field theory duals. 
We showed that our full solution can be viewed as a six-vector deformation away from this AdS${}_3$ limit. 
This deformation was inherited from the interpolation of the original F1-NS5 solution from its AdS${}_3$ limit (in the near horizon region of the F1s) to the asymptotic linear dilaton spacetime associated to the pure NS5 near horizon limit. 
This interpolation has been argued to correspond to a `single-trace' variant of the $T \bar T$ deformation in the CFT${}_2$ dual of the AdS${}_3$ limit \cite{Giveon:2017nie} (the CFT dual (to the long string sector) of string theory on AdS${}_3$ is a symmetric product $\mathcal{M}^{N_1}/S_{N_1}$ and the $T\bar T$ deformation of \cite{Giveon:2017nie} deforms the block CFT $\mathcal{M} \rightarrow \mathcal{M}_{T \bar T}$).

The immediate question is whether there is an analogous interpretation applicable to our six-vector deformation of our AdS${}_3$ limit in terms of a deformation of the CFT duals of \cite{Lozano:2020bxo}.
This is not to necessarily suggest that this deformation will again be describable as a $T \bar T$ deformation, but it may have similar properties.
In general, we would expect generalised U-duality, as for non-abelian T-duality, to produce backgrounds with different CFT duals.
However, we can at least say that our solution generating technique preserved the fact that there \emph{is} a deformation, encoded geometrically, and suggest that this may turn out to have a relationship to $T \bar T$.

A further comment is that in the F1-NS5 case, the endpoint of the deformation can be viewed as a vacuum of the Little String Theory \cite{Aharony:1998ub,Giveon:1999zm} dual to the asymptotic linear dilaton spacetime: for our solution, the latter spacetime maps to the 11-dimensional solution \eqref{dualtoNS5} (not an AdS geometry) which may accordingly itself have a similar holographic interpretation in terms of a dual M5 brane theory.

It may be therefore be interesting to study the deformation of the general class of geometries \eqref{targetAdS} of \cite{Lozano:2020bxo}.
If we define
\be
\begin{split} 
g_{ab} \dd x^a \dd x^b & = \Delta \left(
\frac{u}{\sqrt{\hat{h}_4 h_8}}  r^2 (-\dd t^2+\dd z^2)  
+\frac{\sqrt{\hat{h}_4 h_8}}{u} \dd \varrho^2 \right)
\,,\quad
h_{\alpha \beta} \dd x^{\alpha} \dd x^{\beta}  = \frac{h_8^2}{\Delta^2} \dd s^2_{\text{S}^3/\mathbb{Z}_k} \,,\\
G_{\mu\nu} \dd x^{\mu} \dd x^{\nu} & =\Delta
\left(
\frac{u}{\sqrt{\hat{h}_4 h_8}} \frac{\dd r^2}{r^2} 
+\sqrt{\frac{\hat{h}_4}{h_8}} \dd s_{\text{CY}_2}^2
\right) \,,
\end{split}
\ee
and make the naturally analogous gauge choice
\be
\begin{split} 
C_1 & \equiv C_{t z \rho} = \frac{r^2}{2} \partial_{\varrho} \left(\frac{uu'}{2\hat{h}_4}+2\varrho h_8\right)
\,,\quad C_2 \equiv C_{\psi \theta \phi}  = 2h_8 \left(-\varrho+\frac{u u'}{4\hat{h}_4h_8+u'^2}\right) \sqrt{g_{\text{S}^3/\mathbb{Z}_k}} \,,\\
C_6 &= -\frac{r^2}{2}  \frac{4 h_8^2 u^2 \hat h_4^{\prime 2}}{h (4 h_8 \hat{h}_4 + u'^2)}  \sqrt{g_{\text{S}^3/\mathbb{Z}_k}} + \tfrac{1}{2} C_1 C_2   \,,
\end{split}
\ee
then we can immediately read off a deformed background from the expressions in appendix \ref{e6technology}.
This requires choosing a deformation parameter which produces a new solution: this is not guaranteed.
Note that generically the $\Gsix$ generalised metric block $\gM_{\ii\jj}$ is non-zero for the metric and potentials picked here. This means that the deformed metric will depend quadratically on $\lambda$ instead of just linearly. This is not necessarily a problem, however it is possible that situations with vanishing $\gM_{\ii\jj}$ are special.

Other deformations of the AdS${}_3$ limit of the F1-NS5 solution correspond to single-trace $J\bar T$/$\bar J T$ deformations of the dual CFT${}_2$, see for instance \cite{Chakraborty:2019mdf,Apolo:2019zai}.
These again have a straightforward worldsheet interpretation as TsT i.e. $O(d,d)$ transformations, and modify the bulk geometry.
Focusing on deformations which preserve the ansatz for type IIA on $\text{S}^3$, it would be possible to map the corresponding backgrounds to new 11-dimensional geometries using our methodology, and to examine how the deformations are inherited by the new solution, as trivector deformations for example.

It may also be productive to explore these deformations algebraically in the context of the EDA proposal.
For instance, embedding our $\Gfour$-valued trivector into $\Gsix$ and combining with the six-vector deformation discussed in section \ref{sixvectordeformation}, could be viewed through the lens of the $\Gsix$ EDA \cite{Malek:2020hpo}.
This may connect to related work on polyvector deformations, including in the context of the EDA construction, such as \cite{Gubarev:2020ydf}.

\section*{Acknowledgements}

CB is supported by an FWO-Vlaanderen Postdoctoral Fellowship, and SZ by an FWO-Vlaanderen PhD Fellowship.
The authors further acknowledge the support of the FWO-Vlaanderen through the project G006119N and by the Vrije Universiteit Brussel through the Strategic Research Program ``High-Energy Physics''. 
We would like to thank Marine De Clerck, Saskia Demulder, Camille Eloy and Ondrej Hulik for useful discussions, Carlos Nunez and Daniel Thompson for numerous useful discussions and for providing detailed feedback on drafts of this work, and also the participants of an \href{https://sites.google.com/view/egseminars/past-seminars?authuser=0#h.f9go4iwu2yc2}{Exceptional Geometry online seminar} given by CB for useful questions and comments.

\appendix

\section{Ingredients}
\label{ingredients}

\subsection{Five-brane near horizon limit of pp-F1-NS5}
\label{ppF1NS5} 

\paragraph{Initial solution}
We adapt the notation of \cite{Ortin:2004ms,Chakraborty:2020swe}.
The non-extremal pp-F1-NS5 solution is
\be
\begin{split}
ds_s^2 & = f_1^{-1} ( - f_n^{-1} W \dd t^2 + f_n ( \dd z + \tfrac{1}{2} \tfrac{r_0^2 \sinh 2 \alpha_n}{f_n r^2} \dd t )^2 ) + f_5 ( W^{-1} \dd r^2 + r^2 \dd s^2_{\text{S}^3}) + \dd s^2_{\text{T}^4} \,,\\
B_{tz} & = -\tfrac{1}{2} \tfrac{r_0^2 \sinh 2 \alpha_1}{f_1 r^2}\,,\quad
B_{tz 1 \dots 4}  = - g_s^{-2}  \tfrac{1}{2} \tfrac{r_0^2 \sinh 2 \alpha_5}{f_5 r^2} \,,\quad e^{-2\varphi} = g_s^{-2} f_1 f_5^{-1}\,,
\end{split} 
\label{ppF1NS5clean1}
\ee
where
\be
\begin{split} 
f_n &= 1 + \tfrac{r_n^2}{r^2} \,,\quad
f_1 = 1 + \tfrac{r_1^2}{r^2} \,,\quad
f_5 = 1 + \tfrac{r_5^2}{r^2} \,,\quad
W = 1 - \tfrac{r_0^2}{r^2} \,,
\\
r_1^2& = r_0^2 \sinh^2 \alpha_1 \,,\quad
r_5^2 = r_0^2 \sinh^2 \alpha_5 \,,\quad
r_n^2 = r_0^2 \sinh^2 \alpha_n \,,
\end{split}
\label{ppF1NS5cleanH1}
\ee
and in terms of the numbers $N_1, N_5, N_n$ of strings, five-branes and pp-waves, as well as the (dimensionless) volume parameter $v$ of the $\text{T}^4$, we have
\be
\sinh 2 \alpha_1 = \tfrac{2 N_1 l_s^2}{v} \tfrac{g_s^2}{r_0^2} \,,\quad
\sinh 2 \alpha_5 = \tfrac{2 N_5 l_s^2}{r_0^2} \,,\quad
\sinh 2\alpha_n =  \tfrac{2 N_n l_s^4}{R_x^2 v} \tfrac{ g_s^2}{r_0^2}\,.
\label{ppF1NS5paramcharges}
\ee
The extremal limit sends $r_0 \rightarrow 0$ and $\alpha_1, \alpha_5, \alpha_n \rightarrow \infty$ such that $r_0^2 \sinh 2\alpha_1$, $r_0^2 \sinh 2 \alpha_5$ and $r_0^2 \sinh 2\alpha_n$ are constant and given by \eqref{ppF1NS5paramcharges}.
Then $\sinh \alpha_a^2 \approx \tfrac{1}{2} \sinh 2 \alpha_a$ and so
\be
r_1^2 = \tfrac{N_1 l_s^2 g_s^2}{v}\,,\quad
r_5^2 = N_5 l_s^2 \,,\quad
r_n^2 = \tfrac{N_n l_s^4 g_s^2}{R_x^2 v}
\ee

\paragraph{NS5 near horizon limit}
To obtain a solution we can apply our reduction and uplift procedure to, we need to go to the NS5 near horizon limit.
This limit can be taken by sending the string coupling to zero such that
\be
g_s \rightarrow 0 \,,\quad \tfrac{r_0}{l_s g_s} \,\,\text{fixed} \,.
\ee
This is the Little String Theory (LST) limit \cite{Aharony:1998ub,Giveon:1999zm}.
In this limit, $\alpha_1$ and $\alpha_n$ are fixed, but 
\be
\sinh 2 \alpha_5 \approx \tfrac{2 N_5l_s^2}{r_0^2} \rightarrow \infty
\Rightarrow f_5 \rightarrow \tfrac{N_5 l_s^2}{r^2} \,.
\ee
If we define $u =\frac{r}{l_s g_s}$, $u_0 = \frac{r_0}{l_s g_s}$, then
the three-charge background then becomes in the limit
\be
\begin{split}
ds_s^2 & = f_1^{-1} ( - f_n^{-1} W \dd t^2 + f_n ( \dd z + \tfrac{1}{2} \tfrac{u_0^2 \sinh 2 \alpha_n}{f_n u^2} \dd t )^2 ) + N_5 l_s^2  W^{-1} \frac{du^2}{u^2} +  N_5 l_s^2  \dd s^2_{\text{S}^3} + \dd s^2_{\text{T}^4} \,,\\
H_3 & = - \frac{u_0^2 \sinh 2 \alpha_1}{2} d ( \frac{1}{f_1 u^2} ) \wedge \dd t \wedge \dd x + 2 N_5 l_s^2 \text{Vol}(\text{S}^3) \,,\\
e^{-2\varphi} & = N_5^{-1} u^2 f_1 \,,
\end{split} 
\label{ppF1NS5limitu}
\ee
with
\be
f_1 = 1 + \tfrac{u_0^2 \sinh^2 \alpha_1}{u^2} \,,\quad
f_n = 1 + \tfrac{u_0^2 \sinh^2 \alpha_n}{u^2} \,,\quad
W = 1 - \tfrac{u_0^2}{u^2} 
\,.
\ee
Redefining $u= r^\prime /l_s$, $u_0 = r_0^\prime / l_s$ and immediately \dd ropping the primes we obtain the background in the form \eqref{ppF1NS5limitrfinalcharges}.
In effect this is just the original three-charge background with the $``1+"$ \dd ropped from $f_5$ and $g_s$ set to 1.

\subsection{$\mathrm{CSO}(4,0,1)$ from IIA on $\text{S}^3$}
\label{IIAS3ansatz}

This gauging is known to result from a warped reduction of IIA SUGRA on $\text{S}^3$ \cite{Cvetic:2000dm,Cvetic:2000ah}.
For the pp-F1-NS5 solution, we only need to make use of the NSNS sector reduction ansatz.
Here we need to introduce $\mu^a$, $a=1,\dots,4$ as constrained coordinates on the $\text{S}^3$, $\delta_{ab} \mu^a \mu^b= 1$, a unit determinant symmetric matrix\footnote{Note that what we call $M_{ab}$ is denoted $M^{-1}_{\alpha \beta}$ in \cite{Cvetic:2000ah}.} $M_{ab}$ with inverse $M^{ab}$, and define
\begin{equation}
 U=2M^{ab} M^{bc} \mu^{a}\mu^{c}-\Delta M^{aa}\,, \quad \Delta=M^{ab}\mu^{a}\mu^{b}\,.
\end{equation}
Then the ansatz is
\begin{equation}
\begin{split}
ds_{s}^{2} &=\Phi^{1/2} ds_{7}^{2}+\frac{1}{g^{2}} \Delta^{-1}M_{ab}^{-1}D\mu^{a}D\mu^{b}
\,,\quad
e^{2\varphi} =\Delta^{-1}\Phi^{5/4}\,,\\
H_{3} &= \widetilde F_{(3)}
- \tfrac{1}{2} \epsilon_{a_1 a_2 a_3 a_4}g^{-1}  \Delta^{-1}   F_{(2)}^{a_1 a_2} \wedge D \mu^{a_3} M^{a_4 b} \mu^b  
\\ & \quad
- \tfrac{1}{6}\epsilon_{a_1 a_2 a_3 a_4}g^{-2}  \Delta^{-2} \big(
U \mu^{a_1} D \mu^{a_2} \wedge D \mu^{a_3} \wedge D \mu^{a_4} 
+3 D\mu^{a_1} \wedge D \mu^{a_2}\wedge D M^{a_3 b} M^{a_4 c} \mu^b \mu^c 
\big)\,,
\label{S3ansatz}
\end{split}
\end{equation}
where $D \mu^a \equiv d\mu^a + g A_{(1)}{}^{ab} \mu^b$, $D M^{ab} = d M^{ab} + 2 g A_{(1)}{}^{(a|c} M^{c|b)}$ and $F_{(2)}{}^{ab} = d A_{(1)}{}^{ab} +g A_{(1)}{}^{ac} \wedge A_{(1)}{}^{cb}$.
However these Kaluza-Klein gauge potentials will play no role in the cases we consider.
Although we only write here the ansatz in the NSNS sector, we do need to make use of the full ansatz of \cite{Cvetic:2000ah} to identify the $\Gfour$ covariant multiplets that result.
For instance, the ansatz for the RR four-form field strength introduces a further four three-forms. These combine with the single three-form $\widetilde{F}_{(3)}$ in \eqref{S3ansatz} to form the five-dimensional representation of $\Gfour$. Similarly the scalars $M_{ab}$ and $\Phi$ are joined by four additional scalar fields from the RR sector in order to obtain the full scalar coset $\Gfour/\mathrm{SO}(5)$.
With the RR contribution set to zero, the $\Gfour$ covariant scalar matrix $\gM_{\fA \fB}$, and accompanying scalar potential $V$, are given by:
\be
\gM_{\fA \fB} = \begin{pmatrix} \Phi^{-1/4} M_{ab} & 0 \\ 0 & \Phi \end{pmatrix} \,,
\quad 
V = \tfrac{1}{2} g^2 \Phi^{1/2} ( 2 M^{ab} \delta_{bc} M^{cd} \delta_{ad} - ( M^{ab} \delta_{ab} )^2 )\,.
\label{spot}
\ee

\subsection{Exceptional field theory dictionary}
\label{exft}

Exceptional field theory (see the review \cite{Berman:2020tqn}) describes 11-dimensional supergravity backgrounds after splitting into a $d$-dimensional internal part, with coordinates $x^i$, and $(11-d)$-dimensional external part, with coordinates $X^\mu$.
Fixing the 11-dimensional Lorentz symmetry we write the metric as
\be
ds_{11}^2  = \phi^{-\tfrac{1}{9-d}} g_{\mu\nu} dX^\mu dX^\nu + \phi_{ij} Dx^i Dx^j \,,\quad Dx^i \equiv \dd x^i + A_\mu{}^i dX^\mu \,,
\label{11Decomp}
\ee
where $\phi \equiv \det \phi_{ij}$. 
The three-form and its four-form field strength are decomposed as follows:
\be
 C_{(3)} = \Cdef_{(3)} + \Cdef_{(2) i } Dx^i  + \tfrac{1}{2} \Cdef_{(1) i j } Dx^i Dx^j + \tfrac{1}{3!} \Cdef_{ijk} Dx^i Dx^j Dx^k\,,
\label{Cdecomp_compact}
\ee
\be
  F_{(4)} = \Fdef_{(4)} + \Fdef_{(3) i } Dx^i  + \tfrac{1}{2} \Fdef_{(2) i j } Dx^i Dx^j + \tfrac{1}{3!} \Fdef_{(1) ijk} Dx^i Dx^j Dx^k + \tfrac{1}{4!} \Fdef_{ijkl} Dx^i Dx^j Dx^k Dx^l \,,
\label{Fdecomp_compact}
\ee
where the $(p)$ subscript denotes an $n$-dimensional $p$-form and all wedge products are implicit. 

The fields carrying purely internal indices enter a generalised metric parametrising a coset $\Edd/ H_{d}$, while those carrying external indices (asides from the external metric, $g_{\mu\nu}$) are treated as components of $(11-d)$-dimensional forms in a tensor hierarchy. For instance, one has $\mathcal{A}_\mu{}^M \sim ( A_\mu{}^i, \Cdef_{\mu ij} , \dots )$. Here one has to eventually include components of the dual six-form (and putative dualisations of the metric). In this way, each $p$-form gives a representation of $\Edd$.

For $d=4$, we have ${E}_{4(4)} = \Gfour$. Let $\fM=1,\dots,5$ denote a fundamental index of $\Gfour$.
The generalised metric is represented by a five-by-five unit determinant symmetric matrix: 
\be
\mathcal{M}_{\fM \fN} =
\phi^{\tfrac{1}{10}}
\begin{pmatrix}
\phi^{-\tfrac{1}{2}} \phi_{ij} & - \phi^{-\tfrac{1}{2}} \phi_{i k} C^k \\
- \phi^{-\tfrac{1}{2}} \phi_{i k} C^k  & \phi^{\tfrac{1}{2}} + \phi^{-\tfrac{1}{2}} \phi_{kl} C^k C^l
\end{pmatrix} \,,
\label{SL5genm}
\ee
where $C^i \equiv \tfrac{1}{6} \epsilon^{ijkl} \Cdef_{jkl}$, $\Cdef_{ijk} = - \epsilon_{ijkl} C^l$.
The relevant part of the $\Gfour$ tensor hierarchy consists of gauge fields $\Aa_\mu{}^{\fM \fN} = - \Aa_\mu{}^{\fN \fM}$, $\Ab_{\mu\nu \fM}$, $\Ac_{\mu\nu\rho}{}^{\fM}$, with field strengths $\Fa_{\mu\nu}{}^{\fM\fN}$, $\Fb_{\mu\nu\rho \fM}$, $\Fc_{\mu\nu\rho\sigma}{}^{\fM}$.
These field strengths can be identified with components of the eleven-dimensional four-form and its seven-form dual as follows:
\be
\begin{split} 
\mathcal{F}_{\mu\nu}{}^{i5}& = F_{\mu\nu}{}^i\,,\quad\quad\,\,
\Fa_{\mu\nu}{}^{ij} = \tfrac{1}{2} \epsilon^{ijkl} (\Fdef_{\mu\nu kl} -  \Cdef_{klm} \Fdef_{\mu\nu}{}^{m})\,,\\
\Fb_{\mu\nu\rho i} & = - \Fdef_{\mu\nu\rho i} \,,\quad
\Fb_{\mu\nu\rho 5}  =  \tfrac{1}{4!} \epsilon^{ijkl} (  \rpm \Fdef_{\mu\nu \rho ijkl} +  4  \Fdef_{\mu\nu\rho i}  \Cdef_{jkl})\,,\quad
\\
\Fc_{\mu\nu\rho\sigma}{}^5 & = - \Fdef_{\mu\nu\rho \sigma}\,,\quad\,\,
\Fc_{\mu\nu\rho\sigma}{}^i =  \tfrac{1}{3!} \epsilon^{ijkl}( \rmp \Fdef_{\mu\nu\rho\sigma jkl} - \Cdef_{jkl} \Fdef_{\mu\nu\rho\sigma} )
\label{deftildeF}\,.
\end{split} 
\ee
The bare three-forms appear here as these field strengths transform covariantly under generalised diffeomorphisms.
The minus signs are fixed such that the Bianchi identities of ExFT in the conventions used reproduce those of 11-dimensional supergravity, with $d \hat F_7 \rpm \tfrac{1}{2}\hat F_4 \wedge \hat F_4 = 0$.

\subsection{Exceptional Drinfeld algebra frame}
\label{edaframe}

\paragraph{Generalised frames}

A generalised frame in the $\Gfour$ ExFT can be represented in the 10- or 5-dimensional representations.
However we can only take the generalised Lie derivative with respect to generalised frames $E_{\fA\fB}$ in the former.
The algebra of generalised frames is 
\be
\mathcal{L}_{E_{\fA \fB}} E^{\fM}{}_{\fC} = - F_{\fA\fB \fC}{}^{\fD} E^{\fM}{}_{\fD} \,,
\label{5alg}
\ee
or
\be
 \mathcal{L}_{ E_{\fA \fB}}  E^{\fM \fN}{}_{\fC \fD}
= - \tfrac{1}{2} F_{\fA \fB ,\,\fC \fD}{}^{\fE \fF} E^{\fM \fN}{}_{\fE \fF} 
\,,\quad
F_{\fA \fB,\, \fC \fD}{}^{\fE \fF}
= 4 F_{\fA \fB [\fC}{}^{[\fE} \delta^{\fF]}_{\fD]}\,.
 \label{10alg_appendix}
\ee
The gauging $F_{\fA \fB \fC}{}^{\fD}$ can be decomposed in terms of irreducible representations of $\Gfour$
\be
F_{\fA \fB \fC}{}^{\fD} =  Z_{\fA \fB \fC}{}^{\fD} + \tfrac{1}{2} \delta^{\fD}_{[\fA} S_{\fB] \fC}
 - \tfrac{1}{6} \tau_{\fA \fB} \delta^{\fD}_{\fC} - \tfrac{1}{3} \delta^{\fD}_{[\fA} \tau_{\fB]\fC}\,.
\label{sl5tau}
\ee
Here $\tau_{\fA \fB} \in \mathbf{\overline{10}}$ is the so-called trombone gauging, $S_{\fA \fB} \in \mathbf{\overline{15}}$ and $Z_{\fA \fB \fC}{}^{\fD} \in \mathbf{40}$ obeys $Z_{\fA \fB \fC}{}^{\fD} = Z_{[\fA \fB \fC]}{}^{\fD}$, $Z_{\fA \fB \fC}{}^{\fC} = 0$.

\paragraph{Exceptional Drinfeld algebra frame}
For the exceptional Drinfeld algebra introduced in \cite{Sakatani:2019zrs,Malek:2019xrf} one has
\be
\tilde E^{\fM}{}_{\fA} = \Delta^{-\tfrac{1}{2}} \begin{pmatrix}
l^{\tfrac{1}{2}} \alpha^{\tfrac{1}{2} } v^i{}_a & 0 \\
l^{-\tfrac{1}{2}} \alpha^{-\tfrac{1}{2}} \trivector_a & l^{-\tfrac{1}{2}} \alpha^{-\tfrac{1}{2}} 
\end{pmatrix} \,,\quad
\Delta \equiv \alpha^{\tfrac{3}{5}} l^{\tfrac{1}{5}}\,,
\label{EDAframe}
\ee
in terms of data $(\alpha, l^a{}_i , v^i{}_a, \trivector_a  = \frac{1}{3!} \epsilon_{bcda} \trivector^{bcd})$ describing a particular group manifold with left-invariant frames $l^a{}_i$ and a trivector $\trivector^{abc}$, obeying certain compatibility and differential conditions, in particular
\be
d l^a = \tfrac{1}{2} f_{bc}{}^a l^b \wedge l^c \,,\quad L_{v_a} v_b=-f_{ab}{}^c v_c \,, \quad L_{v_a}\ln \alpha  =\tfrac{1}{3} \frak{L}_a \equiv  \tfrac{1}{3} ( \tau_{a5} - f_{af}{}^f )\,,
\label{dgroup}
\ee
\be
d \trivector^{abc} = \tilde f^{abc}{}_d l^d + 3 f_{ed}{}^{[a} \trivector^{bc]d} l^e + \tfrac{1}{3} \trivector^{abc} \frak{L}_d l^d \,.
\label{dlambda}
\ee
These imply that the components of the gaugings are
\begin{equation}
\begin{aligned}
S_{55} = 0 \,,\quad
Z_{abc}{}^5 = 0 \,,\quad
Z_{ab5}{}^5 = \tfrac{2}{3} \tau_{ab}\,,\quad
Z_{abc}{}^d = -\tau_{[ab} \delta_{c]}^d \, ,
\\
S_{a5} = - \tfrac{2}{3} \tau_{a5} - \tfrac{4}{3} f_{ab}{}^b \,,\quad
Z_{ab5}{}^c = - f_{ab}{}^c - \tfrac{2}{3} \delta_{[a}^c f_{b]d}{}^d \,.
\label{edaflux2}
\end{aligned} 
\end{equation}
while $S_{ab}$ and $\tau_{ab}$ are defined via the ``dual'' structure constant with three upper antisymmetric indices  
\be
\tilde f^{abc}{}_d =   \frac{1}{4}  \epsilon^{abce} ( S_{de} + 2 \tau_{de} ) \,.
\label{fStau}
\ee  
In terms of generators $T_{\fA \fB}$ obeying $[T_{\fA\fB}, T_{\fC\fD}] = \tfrac{1}{2} F_{\fA\fB, \fC\fD}{}^{\fE\fF} T_{\fE\fF}$ the algebra can be written in a compact form reminiscent of the Drinfeld double algebra if we let $T_a \equiv T_{a5}$, $\tilde T^{ab} \equiv \tfrac{1}{2} \epsilon^{abcd} T_{cd}$. The brackets are:
\be
\begin{split}
[T_a, T_b ] & =  f_{ab}{}^c T_c \,, \qquad [\tilde T^{ab}, \tilde T^{cd} ]  =   2 \tilde f^{ab[c}{}_e \tilde T^{d]e}   \,,\\
[T_a, \tilde T^{bc} ] &  =    2  f_{ad}{}^{[b} \tilde T^{c]d} - \tilde f^{bcd}{}_a T_d - \tfrac{1}{3} \frak{L}_a  \tilde T^{bc}   \, ,   \quad
 [\tilde T^{bc} , T_a ]  =     3 f_{[de}{}^{[b} \delta_{a]}^{c]} \tilde T^{de} 
  + \tilde  f^{bcd}{}_a T_d 
  +  \frak{L}_d  \delta_a^{[b} \tilde T^{cd]} 
   \,.
\end{split}
\label{EDAf}
\ee

\paragraph{$\mathrm{CSO}(4,0,1)$ frame and scalar potential} 

This frame has $\alpha=1$, $v^i{}_a =\delta^i_a$ and $\trivector^{abc} = g \epsilon^{abcd} x_d$ \cite{Blair:2020ndg}  (where we use $\delta^i_a$ to identify the curved and flat indices on $x^i$ and $\delta_{ab}$ to raise/lower).
This results in $\tilde f^{abc}{}_d = g \epsilon^{abc}{}_d$ or equivalently $S_{ab} = 4 g \delta_{ab}$, with the other structure constants components all vanishing.

When $S_{\fA \fB} \neq 0$ is the only non-vanishing $\Gfour$ gauging, the scalar potential resulting from ExFT is in our conventions
\be
V = \frac{1}{32} \left( 2 \gM^{\fA \fB} S_{\fB \fC} \gM^{\fC \fD} S_{\fD \fA} - ( \gM^{\fA \fB} S_{\fA \fB} )^2\right)\,.
\ee
For the $\mathrm{CSO}(4,0,1)$ case with the scalar matrix as in \eqref{spot} and the gauging $S_{\fA \fB}$ resulting from the EDA frame, this exactly matches the scalar potential of \eqref{spot}.

\subsection{$\Gsix$ generalised metric for a 3+3 split and six-vector deformation}
\label{e6technology}

\paragraph{Components} 

Write the six-dimensional index as $\ii = ( a,\alpha)$, where both $a$ and $\alpha$ are three-dimensional.
Consider the case where
\be
\phi_{\ii\jj} = \begin{pmatrix} g_{ab} & 0 \\ 0 & h_{\alpha \beta} \end{pmatrix} \,,\quad
C_{\ii\jj\kk} \rightarrow ( C_1 \epsilon_{abc}, C_2 \epsilon_{\alpha \beta \gamma}) \,,\quad
\epsilon_{abc\alpha\beta\gamma} = \epsilon_{abc} \epsilon_{\alpha \beta \gamma} \,,
\ee
and $C_{\ii_1 \dots \ii_6} = C_6 \epsilon_{\ii_1 \dots \ii_6}$.
Let $t$ denote the number of timelike directions of the metric $\phi_{\ii\jj}$, and let $g \equiv \det (g_{ab})$, $h \equiv \det(h_{\alpha \beta})$.
The components of the $\Gsix$ generalised metric defined by \eqref{defE6GM} can then be computed block-by-block to be
\be
\begin{split} 
\gM_{ab} & = |\phi|^{1/3}  g_{ab} \left( 1+ \tfrac{1}{gh} ( h C_1^2 + (C_6 + \tfrac{1}{2} C_1 C_2 )^2 )\right) \,,\\
\gM_{\alpha \beta}  & = |\phi|^{1/3}  h_{\alpha\beta} \left( 1+ \tfrac{1}{gh} ( g C_2^2 + (C_6 - \tfrac{1}{2} C_1 C_2 )^2 )\right) \,,\quad \gM_{a \alpha}  = 0\,, \,
\end{split} 
\ee
\be
\begin{split}
 \gM_{a}{}^{bc} & = - (-1)^t |\phi|^{-2/3}  g_{ad} \epsilon^{bcd} ( h C_1 + C_2 (C_6 + \tfrac{1}{2} C_1 C_2 ) ) \,,\\
 \gM_{\alpha}{}^{\beta\gamma} & = - (-1)^t |\phi|^{-2/3}  h_{\alpha \delta} \epsilon^{\beta\gamma\delta} ( gC_2 - C_1 (C_6 - \tfrac{1}{2} C_1 C_2 ) ) \,,\\
\gM_{a}{}^{\beta\gamma} &= 0 = \gM_{\alpha}{}^{bc} = \gM_{b}{}^{a \alpha} = \gM_{\beta}{}^{a \alpha} \,,
\end{split}
\ee
\be
\begin{split} 
\gM_{a\bar b} & = (-1)^t |\phi|^{-2/3}  g_{ab}  (C_6 + \tfrac{1}{2} C_1 C_2 ) \,,\quad 
\gM_{\alpha \bar \beta} =(-1)^t |\phi|^{-2/3}    h_{\alpha\beta} (C_6 - \tfrac{1}{2} C_1 C_2 ) \,,\\
\gM_{a\bar \alpha} & =  \gM_{\alpha \bar a}=0\,,
\end{split} 
\ee
\be
\begin{split}
 \gM^{ab}{}_{\bar c}& =- (-1)^t |\phi|^{-2/3}  g_{cd} \epsilon^{dab} C_2  \,,\quad
 \gM^{\alpha\beta}{}_{\bar \gamma}  =  (-1)^t |\phi|^{1/3}  h_{\gamma\delta} \epsilon^{\delta \alpha \beta} C_1 \,,\\
\gM^{a \alpha}{}_{\bar b}  &  = \gM^{a \alpha}{}_{\bar \beta} = \gM^{ab}{}_{\bar \alpha} 
= \gM^{\alpha \beta}{}_{\bar a} = 0
\end{split} 
\ee
\be
\begin{split}
\gM_{\bar a \bar b} & =  (-1)^t |\phi|^{-1/3}  g_{ab} \,,\quad
\gM_{\bar \alpha \bar \beta}  = (-1)^t |\phi|^{1/3}  h_{\alpha \beta} \,,\quad
\gM_{\bar a \bar \alpha}  =0 \,,\\
\end{split}
\ee

\paragraph{Six-vector deformation} 

Using \eqref{defUOmega}, one sees that the six-vector deformation has the relatively simple effect of
\be
\gM_{\ii \bar \jj} \rightarrow \gM_{\ii \bar \jj} + \tilde \Omega \gM_{\ii \jj} \,,\quad
\gM^{\ii\ii'}{}_{\bar \jj} \rightarrow \gM^{\ii\ii'}{}_{\bar \jj} + \tilde \Omega \gM^{\ii\ii'}{}_{\jj} \,,\quad
\gM_{\bar \ii \bar \jj} \rightarrow \gM_{\bar \ii \bar \jj} + \tilde \Omega (\gM_{\ii \bar \jj} + \gM_{\jj \bar \ii} ) + \tilde \Omega^2 \gM_{\ii \jj} 
\ee
leaving other blocks invariant.
Then given a configuration with
\be
ds^2_{11} = g_{ab} \dd x^a \dd x^b + h_{\alpha \beta} \dd x^\alpha \dd x^\beta + G_{\mu\nu} \dd x^\mu \dd x^\nu 
\ee
and gauge field components $C_1$ and $C_2$ and $C_6$ as above, the effect of a six-vector deformation is to produce the following metric and gauge fields:
\be
\begin{split} 
\widetilde{ds^2_{11}} & = (1 + \Theta_1)^{1/3} (1+\Theta_2)^{-2/3} g_{ab} \dd x^a \dd x^b
 +  (1 + \Theta_1)^{-2/3} (1+\Theta_2)^{1/3} h_{\alpha\beta} \dd x^\alpha \dd x^\beta
 \\ & \quad
 +  (1 + \Theta_1)^{1/3} (1+\Theta_2)^{1/3} G_{\mu\nu} \dd x^\mu \dd x^\nu \,,
\end{split}
\ee
\be
\begin{split} 
\widetilde{C}_1 & =\frac{1}{1+\Theta_2} \left( C_1 - \Omega ( g C_2 - C_1 ( C_6 - \tfrac{1}{2} C_1 C_2 ))\right) \,,\\
\widetilde{C}_2 & =\frac{1}{1+\Theta_1} \left( C_2 + \Omega ( h C_1 + C_2 ( C_6 + \tfrac{1}{2} C_1 C_2 ))\right) \,,\\
\tilde{C}_6 & = \frac{1}{2} \frac{1}{1+\Theta_1} ( C_6 + \tfrac{1}{2} C_1 C_2 + \Omega (gh + h C_1^2 + (C_6 + \tfrac{1}{2} C_1 C_2)^2)) 
\\ & \qquad
+ \frac{1}{2} \frac{1}{1+\Theta_2} ( C_6 -\tfrac{1}{2} C_1 C_2 + \Omega (gh + g C_2^2 + (C_6 - \tfrac{1}{2} C_1 C_2)^2)) 
\end{split}  
\ee
where
\be
\begin{split}
\Theta_1 & = 2 \Omega ( C_6 + \tfrac{1}{2} C_1 C_2 ) + \Omega^2 (gh + h C_1^2 + (C_6 + \tfrac{1}{2} C_1 C_2)^2) \,,\\
\Theta_2 & = 2 \Omega ( C_6 - \tfrac{1}{2} C_1 C_2 ) + \Omega^2 (gh + g C_2^2 + (C_6 - \tfrac{1}{2} C_1 C_2)^2) \,.
\end{split}
\ee

\section{Charge quantisation}
\label{charges}

In this appendix we consider the requirement of brane charge quantisation for our new solution.
We therefore reinstate the dimensionful constants $r_1$ and $R$ inherited from the original F1-NS5 solution.
We also note that we can include a constant $\alpha$ (assumed dimensionless) in the EDA frame corresponding to the trombone rescaling of the 11-dimensional solution.
Including this, the extremal solution in spherical coordinates would be:
\be
\begin{split} 
\dd s_{11}^2  &=\alpha^{2/3} (r^2 f_1 + \rho^2)^{1/3} R^{-4/3} (r^2 f_1)^{1/3} 
\left(  
\frac{1}{f_1} ( - \dd t^2 + \dd z^2 ) + \frac{R^2 \dd \rho^2}{r^2 f_1} 
+    \frac{R^2 \dd r^2}{r^2} + \dd s^2_{\text{T}^4}
\right)
\\ & \quad
 + \alpha^{2/3}( r^2 f_1 +  \rho^2)^{-2/3} R^{2/3} (r^2f_1)^{1/3} \rho^2 \dd s^2_{\text{S}^3} 
\,,
\\
F_{(4)} & = 
\alpha \frac{2 r_1^2}{(r^2 f_1)^2} \frac{r  \rho}{R} \dd t \wedge \dd z \wedge \dd r \wedge \dd\rho 
\rpm \alpha \frac{2 r_1^2}{R^3} \mathrm{Vol}_{\text{T}^4}
\\ & \qquad + \alpha \frac{R( 4 r^2 f_1 + 2  \rho^2 ) }{(r^2 f_1 +  \rho^2 )^2}  \rho^3 \dd\rho \wedge \mathrm{Vol}_{\text{S}^3} 
- \alpha \frac{R \rho^4}{(r^2 f_1 +  \rho^2)^2}  \partial_r ( r^2 f_1) \dd r \wedge \mathrm{Vol}_{\text{S}^3}  \,.
\end{split} 
\label{solnSphericalalpha}
\ee 
The dual field strength is
\be
\begin{split}
\star F_{(4)} & = 
-\alpha^{2}
\frac{2 r_1^2}{r^2 f_1 +  \rho^2 } \frac{ \rho^4}{ R^2} \mathrm{Vol}_{\text{S}^3} \wedge \mathrm{Vol}_{\text{T}^4}  
 \rpm \alpha^{2}
\frac{2 r_1^2}{r f_1 ( r^2 f_1 + \rho^2 )} \rho^3 \dd t \wedge \dd z \wedge \dd r \wedge \dd \rho \wedge  \mathrm{Vol}_{\text{S}^3} 
\\ & \quad + \alpha^{2}
\frac{2r}{R^4 } ( 2 r^2 f_1 +  \rho^2) \dd t \wedge \dd z \wedge \dd r \wedge\mathrm{Vol}_{\text{T}^4} 
+ \alpha^{2}
\frac{r\rho}{R^4f_1} \partial_r(r^2 f_1) \dd t \wedge \dd z \wedge \dd \rho  \wedge \mathrm{Vol}_{\text{T}^4}\,.
\end{split} 
\ee 
The number of membranes and fivebranes will be determined by
\be
N_{M2} = \tfrac{1}{(2\pi)^6 l_p^6} \int J_{\text{Page}} \,,\quad
N_{M5} = \tfrac{1}{(2\pi)^3 l_p^3} \int F_{(4)} 
\ee
As discussed in section \ref{solnsph}, $J_{\text{Page}}$ vanishes up to large gauge transformations of the form $C_{(3)} \rightarrow C_{(3)} + 4\pi j l_p^3 \mathrm{Vol}_{\text{S}^3}$, $j \in \mathbb{Z}$, which shift $J_{\text{Page}} \rightarrow J_{\text{Page}} +  4\pi j l_p^3 \alpha \tfrac{2 r_1^2}{R^3} \mathrm{Vol}_{\text{S}^3} \wedge \mathrm{Vol}_{\text{T}^4}$.
Hence
\be
N_{M2}
= N_1 4 \pi j \tfrac{l_s^6}{l_p^3} \tfrac{\alpha}{R^3}   \,.
\ee
Now consider the M5 branes.
Integrating the flux through the torus we have
\be
N_{M5}= \rpm  \tfrac{1}{(2\pi)^3 l_p^3}\alpha \tfrac{2r_1^2}{R^3} (2 \pi)^4 v l_s^4 
=\rpm 4\pi N_1  \tfrac{l_s^6}{l_p^3}  \tfrac{\alpha}{R^3} \,.
\ee
Notice that $N_{M2} = j |N^{(\text{T}^4)}_{M5}|$.

Next integrating the flux through the four-cycle in $(r,\rho, \text{S}^3)$ directions as described in section \ref{solnsph} gives, if $r_1=0$
\be
N_{M5{}'}= \tfrac{1}{(2\pi)^3 l_p^3} 2 \pi^2 \alpha R \bar{\rho}^2
 = \tfrac{\alpha R}{4\pi l_p^3} \bar\rho^2
\ee
where $\bar\rho$ corresponds to the limit of the $\rho$ integration (starting at $\rho=0$).
Then charge quantisation requires
\be
\bar\rho^2 = N \tfrac{4 \pi l_p^3}{\alpha R} \,,\quad N \in \mathbb{N} \,.
\ee
The above results work remarkably well with the matching to the AdS solutions of \cite{Lozano:2020bxo}.
Restoring the Planck length appropriately in the solution \eqref{targetAdS} such that $\rho$ has units of length and $\varrho$ is dimensionless, and carefully working through the identification with the AdS limit $r^2f_1 =r_1^2$ of \eqref{solnSphericalalpha}, the matching condition \eqref{match1} and \eqref{match2} become
\be
\rho^2 = \frac{2l_p^3}{R \alpha}\varrho \,,\quad
u = \alpha \frac{2 r_1^2 \varrho}{l_p R} \,,\quad
\hat h_4 = \alpha \frac{2 r_1^2 l_p \varrho}{R^3} \,.
\label{matchalpha}
\ee
In \cite{Lozano:2020bxo} we have a sequence of intervals $\varrho \in [2\pi j, 2\pi (j+1)]$. 
Viewing our solution as lying in the first interval, $\varrho \in [0,2\pi]$ we have $\bar\rho^2 = \tfrac{4\pi l_p^3}{\alpha R}$ giving one unit of charge.
Meanwhile the relationship between the M2 and M5 charges matches that following from equations (3.6) to (3.8) of \cite{Lozano:2020bxo}.

Finally we can try to fix the relationship between the 11-dimensional Planck length and the 10-dimensional string length appearing in the original solutions in type IIA on $\text{S}^3$. 
A crude way to do this is to reduce the Newton's constant prefactor of 11- and 10-dimensional supergravity to the 7-dimensional theory, via
\be
\frac{1}{2\kappa_{11}^2} \int \dd \rho \,\rho^3 \dd \Omega_{3} = 
\frac{1}{2\kappa_{10}^2} \int R^3 \dd\Omega_{3} \Rightarrow
\frac{ 2\pi^2 \tfrac{1}{4} \bar\rho^4}{(2\pi)^8 l_p^9} = \frac{2\pi^2 R^3}{(2\pi)^7 l_s^8} 
\Rightarrow \frac{l_s^3}{l_p^3} = \frac{\alpha^2 N_5^{5/2}}{2\pi} \,,
\ee
which implies 
\be
N_{M5} = {2 N_5 \alpha^3} N_1 \,.
\ee
It seems most natural to take $\alpha=(2N_5)^{-1/3}$, as the field strength component giving rise to this flux comes directly from the three-form flux due to the F1 in the original brane solution.

\bibliography{CurrentBib}

\end{document}